\documentclass[conference]{IEEEtran}
\usepackage{helvet}       % Helvetica as sans-serif font
\usepackage{courier}      % Courier as typewriter font
\usepackage{microtype}
\usepackage{cite}
\usepackage{hyperref}
\usepackage[anythingbreaks]{breakurl}
\usepackage{graphicx}

\usepackage{algorithm}
\usepackage{listings}
\usepackage{soul}
\usepackage{inconsolata}
\usepackage{xspace}
\usepackage{multirow}
\usepackage{adjustbox,lipsum}
\usepackage[font={small,it},labelfont=bf]{caption}

\DeclareGraphicsExtensions{.pdf,.jpeg,.png}
\graphicspath{ {./figs/} }
\usepackage{array}
\usepackage[usenames,dvipsnames]{color}

\newcommand{\tool}{{\sc Robovic}\xspace}

\usepackage{xcolor,colortbl}
\definecolor{Gray}{gray}{0.85}

\newfloat{codelist}{tbp}{loa}
\floatname{codelist}{Listing}

\definecolor{dkgreen}{rgb}{0,.6,0}
\definecolor{dkblue}{rgb}{0,0,.6}
\definecolor{dkyellow}{cmyk}{0,0,.8,.3}
\definecolor{lightlightgray}{rgb}{.98,.98,.98}
\definecolor{lightgray}{rgb}{.93,.93,.93}
\definecolor{gray}{rgb}{.4,.4,.4}
\definecolor{purple}{rgb}{0.65, 0.12, 0.82}
\definecolor{easyred}{rgb}{0.65, 0.12, 0.12}
\definecolor{green}{rgb}{0.12, 0.65, 0.12}
\definecolor{darkgray}{rgb}{0.2, 0.2, 0.2}

\lstdefinelanguage{JavaScript}{
  keywords={typeof, new, true, false, catch, function, return, null, catch, switch, var, if, in, while, do, else, case, break},
  keywordstyle=\color{blue}\bfseries,
  ndkeywords={class, export, boolean, throw, implements, import, this},
  ndkeywordstyle=\color{gray}\bfseries,
  identifierstyle=\color{black},
  sensitive=false,
  comment=[l]{//},
  morecomment=[s]{/*}{*/},
  commentstyle=\color{darkgray}\ttfamily,
  stringstyle=\color{easyred}\ttfamily,
  morestring=[b]',
  morestring=[b]"
}

\lstdefinelanguage{English}{
  keywords={},
  keywordstyle=\color{blue}\bfseries,
  ndkeywords={},
  ndkeywordstyle=\color{gray}\bfseries,
  identifierstyle=\color{black},
  sensitive=false,
  comment=[l]{//},
  morecomment=[s]{/*}{*/},
  commentstyle=\color{darkgray}\ttfamily,
  stringstyle=\color{easyred}\ttfamily,
}

\hypersetup{
  bookmarks=false, % don't show bookmarks tab on startup
  colorlinks=true,%
  citecolor=black,%
  filecolor=black,%
  linkcolor=black,%
  urlcolor=black,%
  pagebackref=true,%
}

\makeatletter
\def\@copyrightspace{\relax}
\makeatother

\begin{document}


\author{\IEEEauthorblockN{Anonymous submission}}

\title{Dial One for Scam:\\A Large-Scale Analysis of Technical Support Scams}

\author{
\IEEEauthorblockN{Najmeh Miramirkhani}
\IEEEauthorblockA{Stony Brook University\\
nmiramirkhani@cs.stonybrook.edu}
\and

\IEEEauthorblockN{Oleksii Starov}
\IEEEauthorblockA{Stony Brook University\\
ostarov@cs.stonybrook.edu}
\and
\IEEEauthorblockN{Nick Nikiforakis}
\IEEEauthorblockA{Stony Brook University\\
nick@cs.stonybrook.edu}
}

\IEEEoverridecommandlockouts
\makeatletter\def\@IEEEpubidpullup{9\baselineskip}\makeatother

%\CopyrightYear{2016}
%\setcopyright{acmlicensed}

\maketitle
%\subsection*{Abstract}
\begin{abstract}
In technical support scams, cybercriminals attempt to convince users that their machines are infected with malware and are in need of their technical support. In this process, the victims are asked to provide scammers with remote access to their machines, who will then ``diagnose the problem'', before offering their support services which typically cost hundreds of dollars. Despite their conceptual simplicity, technical support scams are responsible for yearly losses of tens of millions of dollars from everyday users of the web.

In this paper, we report on the first systematic study of technical support scams and the call centers hidden behind them. We identify malvertising as a major culprit for exposing users to technical support scams and use it to build an automated system capable of discovering, on a weekly basis, hundreds of phone numbers and domains operated by scammers. By allowing our system to run for more than 8 months we collect a large corpus of technical support scams and use it to provide insights on their prevalence, the abused infrastructure, the illicit profits, and the current evasion attempts of scammers. Finally, by setting up a controlled, IRB-approved, experiment where we interact with 60 different scammers, we experience first-hand their social engineering tactics, while collecting detailed statistics of the entire process. We explain how our findings can be used by law-enforcing agencies and propose technical and educational countermeasures for helping users avoid being victimized by technical support scams.
\end{abstract}

\vspace{-3ex}
\section{Introduction}

A recent and understudied social engineering attack targeting everyday web users is a \emph{technical support scam}. In a technical support scam, a webpage created by the scammer tries to convince users that their machines are infected with malware and instructs them to call a technical support center for help with their infection. The victimized users will then willingly provide remote access to their machine and, if the scammer successfully convinces them that they are indeed infected, pay the scammer a malware-removal fee in the range of hundreds of dollars. This scam has become so prevalent that the Internet Crime Complaint Center released a Public Service Announcement in November 2014 warning users about technical support scams~\cite{psa-ic3}.

Even though this type of scam costs users millions of dollars on a yearly basis~\cite{scam-cost,psa-ic3}, there has been no systematic study of technical support scams from the security community. Thus, while today we know that these scams do in fact take place and that scammers are successfully defrauding users, any details about their operations are collected in an unsystematic way, e.g., by victimized users recalling their experiences, and antivirus companies analyzing a handful of scams in an ad-hoc fashion~\cite{malware-bytes-1,malware-bytes-2,mypchas}.

In this paper, we perform a three-pronged analysis of the increasingly serious problem of technical support scams. First, we build a reliable,  distributed crawling infrastructure that can identify technical support scam pages and use it to collect technical support scam pages from websites known to participate in malvertising activities. By deploying this infrastructure, in a period of 250 days, we discover 8,698 unique domain names involved in technical support scams, claiming that users are infected and urging them to call one of the 1,581 collected phone numbers. To the best of our knowledge, our system is the first one that can automatically discover hundreds of domains and numbers belonging to technical support scammers every week, without relying on manual labor or crowdsourcing, which appear to be the main methods of collecting instances of technical support scams used by the industry~\cite{malware-bytes-1,malware-bytes-2}.

Second, we analyze the corpus of collected data and find multiple patterns and trends about the techniques used and the infrastructure abused by scammers. Among others, we find that scammers register thousands of low-cost domain names, such as, \texttt{.xyz} and \texttt{.space}, which abuse trademarks of large software companies and, in addition, abuse CDNs as a means of obtaining free hosting for their scams. We trace the collected phone numbers and find that while 15 different telecommunication providers are abused, four of them are responsible for more than 90\% of the numbers. We show that scammers are actively evading dynamic-analysis systems located on public clouds and find that, even though the average lifetime of a scam URL is approximately 11 days, 43\% of the domains were only pointing to scams for less than 3 days. Moreover, we identify potential campaigns of technical support scams, their unique characteristics, and estimate their life time finding that 69\% of scam campaigns have a lifetime of less than 50 days, yet some survive for the entire duration of our experiment. From a financial perspective, we take advantage of publicly exposed webserver analytics and estimate that, just for a small fraction of the monitored domains, scammers are likely to have made more than 9 million dollars.

Third, we obtain permission from our IRB to conduct 60 sessions with technical support scammers, where we call the numbers discovered by our distributed infrastructure and give scammers access to disposable virtual machines, while recording the entire session. By analyzing the collected data, we calculate precise statistics about the abused tools, the utilized social engineering techniques, and the requested charges. Among others, we find that scammers are patient (average call duration is 17 minutes), abuse a limited number of remote administration tools (81\% of all scammers used one of two software tools), charge victims hundreds of dollars (average charge is \$290.9), and are creative in their ways of convincing users that their machines are infected with a virus (more than 12 different techniques utilized). Moreover, we use a large number of volunteers to estimate the size of call centers operated by scammers and find that the average call center is housing 11 technical support scammers, ready to receive calls from victims.

Finally, we explain why educating the general public about technical support scams should be easier than educating them about other security issues, and propose the development of an, in-browser, ``panic button'' that non-technical users would be educated to use when they feel threatened by the content of any given webpage.

\noindent Our main contributions are:
\begin{itemize}
\itemsep0em 

\item We design and develop the first system capable of automatically discovering domains and phone numbers operated by technical support scammers.
\item We perform the first systematic analysis of technical support scam pages and identify their techniques, abused infrastructure, and campaigns.
\item We interact in a controlled fashion with 60 scammers and collect intelligence that can be used for takedowns, technical countermeasures, and public education.
\item We make a series of propositions for educating the public about technical support scams and for protecting users from abusive pages.

\end{itemize}

\section{Background}
\label{sec:background}

\begin{figure}[t]
\centering
\includegraphics[scale=0.26]{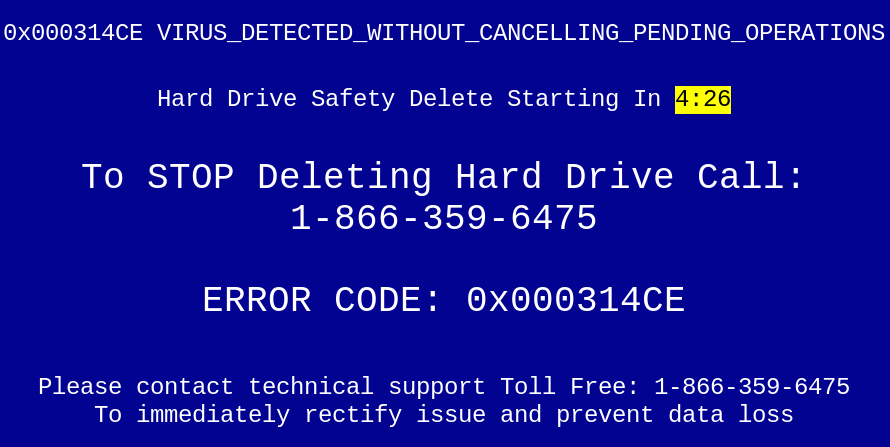}
\caption{Screenshot of a technical support scam which mimics a Windows blue-screen-of-death to increase its trustworthiness.}
\vspace{-5ex}
\label{fig:scam}
\end{figure}

A technical support scam begins with a user landing on a page claiming that her operating system is infected with malware. 
Pages hosting technical support scams typically abuse logos and trademarks of popular software and security companies, or operating system UIs, to increase their trustworthiness. Figure~\ref{fig:scam} shows an example scam page. Instead of requesting from users to download software, as typical scareware scams did in the past~\cite{stone2013underground}, these scams request from the users to call a support center for help with their infection. The posted number is often a toll-free number which is clearly used to increase the chances that a user would actually dial it. Finally, the page is using intrusive JavaScript techniques in order to make it hard for the user to navigate away, such as, constantly showing \texttt{alert} boxes that ask the user to call the technical support number, and hooking into the \texttt{onunload} event, which is triggered when the user attempts to close the current browser tab, or navigate away from the current website.

Once a user calls the listed number, she will eventually reach a person requesting access to her machine in order to diagnose the problem. The user is instructed to download remote desktop software and allow the remote ``technician'' to connect to her machine. After connecting, the scammer, unfortunately, has full control over the user's machine. The scammer will then proceed to demonstrate the infection by showing errors and supposed problems that, in reality, are typical of any Windows installation. As soon as the scammer realizes that the user is convinced, he will then offer to fix the problem for a fee, typically in the range of hundreds of dollars which the user is asked to pay by giving her credit card number to the scammer. As one can clearly understand, the above scenario will, at best, result in the user paying hundreds of dollars for unnecessary services. At worst, the scammer can keep charging the credit card until the limit is reached, install malware and keystroke loggers on the user's machine, and use them to exfiltrate the user's private and financial information~\cite{shuang2015mules}.

\section{Data Collection and Analysis}
\label{sec:robovic}

\subsection{Source of technical support scam pages}

Even though technical support scams are a known phenomenon, the exact details of how a user ends up on a technical support scam page are less known. In order to study this phenomenon, we need access to a steady stream of URLs with high toxicity, similar to the needs of dynamic analysis frameworks for the detection of drive-by downloads~\cite{invernizzi2012evilseed}. 

We argue that most users are exposed to malicious content via malvertising. The constant stream of news of malvertising detected on popular websites~\cite{malvertising1,malvertising2,malvertising3}, and the constant crackdown (and promises of crackdown) from advertising networks~\cite{antimalvertising1,antimalvertising2,antimalvertising3} make this clear. Therefore, even though a scammer could, in theory, try to lure individual users to click on direct links towards his scam pages, this behavior will not only result in a reduced number of victims but also in the faster identification and thus takedown of the malicious page. The natural non-determinism of advertising networks and the ability to trace the provenance of the current visitor, provide ample opportunities for scammers to reveal themselves to victims while hiding from search engines and security researchers.

In this paper, we take advantage of the results of specific recent studies which found that two types of services, namely, \emph{domain parking} and \emph{ad-based URL shorteners}, engage in malvertising practices that endanger users.

\vspace{0.5ex}
\noindent\textbf{Domain Parking.} Domain parking companies compile portfolios of tens of thousands of underdeveloped domain names which they use to show ads to the landing users. If a user clicks on an ad, the domain parking company will then, presumably, give a portion of the advertising profit to the owner of the unused domain. Apart from hiding advertising profits from the domain owners~\cite{Alrwais2014}, many domain parking companies have been found to collaborate with dubious advertising networks which do not hesitate to occasionally redirect a user to a page with malware. In fact, Vissers et al., while researching the types of ads that users who land on parked websites are exposed to, discovered two pages which fit our definition of a technical support scam~\cite{parking_ndss2015}. To find a sufficient number of parked domains that our crawlers can visit, we take advantage of the fact that prior research has shown that domain parking is the favorite monetization method of domain squatters~\cite{edelman2003,Wang:2006:STD:1251296.1251301,Moore2010,Szurdi:long-taile-of-typosquatting,Agten2015seven,khan2015every}. Therefore, as long as we visit typosquatting variants of popular domain names, such as \texttt{twwitter.com} (note the duplication of the \texttt{``w''} character), the majority of our visits will end up on domain parking companies which will redirect a fraction of these visits to technical support scams.

\vspace{0.5ex}
\noindent\textbf{Ad-based URL Shorteners.} Ad-based URL shorteners are services that allow the users who shorten URLs to make a commission every time that other users visit their shortened URLs. Instead of immediately redirecting the short-URL-visiting users, ad-based URL shorteners force users to view an ad for a few seconds, before they can proceed to the intended, ``long'', URL. Nikiforakis et al. studied the ecosystem of ad-based URL shortening services and their ad-delivery methods~\cite{nikiforakis2014stranger}, finding a large percentage of malvertising.

\vspace{0.5ex}
\noindent\textbf{Generality of our approach.} Note that we are not claiming that scammers explicitly collaborate with either domain-parking agencies, or ad-based URL shorteners. Instead, we use these two services as our gateway to malicious advertising, rather than as a method for identifying specific advertisers. As such, we argue that our methodology will be able to detect, with equal probability, all scammers that are using advertising as a way of attracting victims.

\subsection{Tool design and implementation}

\label{sec:tooldesign}

Our tool for discovering and recording technical support scams is called \tool (Robotic Victim). Our main objective is to collect as much data as possible about this highly profitable underground business, in order to conduct a systematic study of technical support scams and analyze their unique characteristics. At the same time, a necessary condition for gathering technical-support-related data is the development of a reliable and highly available infrastructure, that will provide us with enough uptime to be able to study temporal properties of technical support scams. 
Figure~\ref{fig:scam-eco} shows the high-level view of \tool, the high-toxicity, input streams of URLs, and the interactions of our tool with the technical support scam ecosystem. We describe \tool's core components below:

\begin{figure}
\centering
    \includegraphics[scale=0.40]{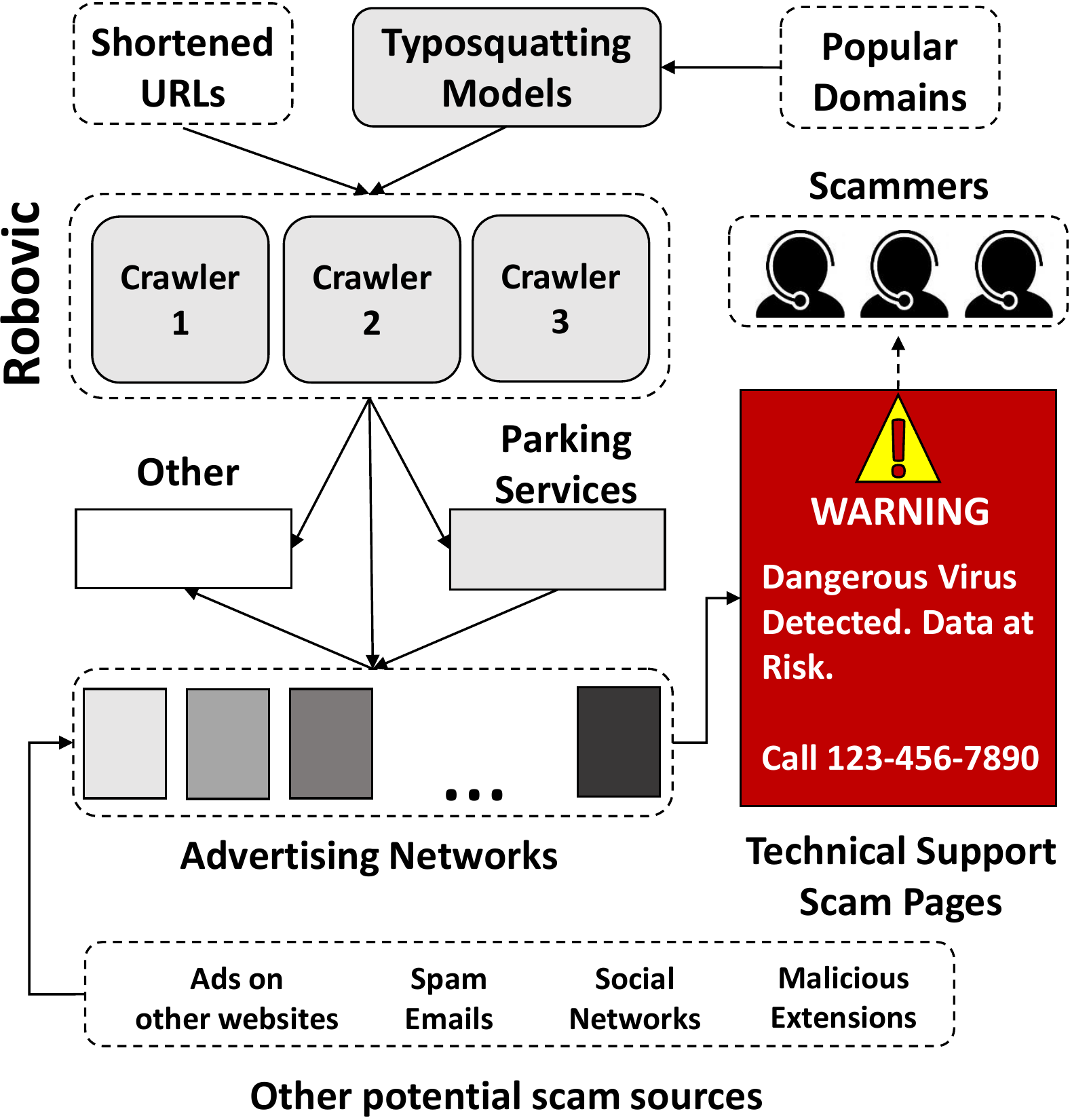}
    \caption{High level view of our automated detection and collection tool of technical support scams (\tool) and its interaction with the technical support scam ecosystem}
    \label{fig:scam-eco}
\vspace{-5ex}
\end{figure}

\vspace{1ex}
\noindent\textbf{Crawler.} The Crawler is in charge of browsing and collecting data for the given set of URLs and recording information about the resulting pages. To address the requirements of our study, we extended OpenWPM which is a generic web privacy measurement platform~\cite{englehardt2015openwpm}. More specifically, we implemented a custom browser extension to instrument the browser so that specific native JavaScript functions, like the aforementioned \texttt{alert} function, would be overwritten before loading a page, in a way that allows us to record the frequency of calls and exact messages displayed to users. In addition, our browser extension ensured the modification of the browser's user-agent properties to match a typical user browsing the web using a Microsoft Windows OS. \tool uses a MITM proxy to record requests and responses, clicks on pop ups and logs the HTML code of all the nested iframes, the final URL, the text shown in alert boxes and the functions used in commonly abused event handlers, such as, the \texttt{onunload} handler, as well as a screenshot of the page. Finally, given the adversarial nature of technical support scam pages, e.g., the locking of a user's browser via the use of the JavaScript-accessible, browser-provided \texttt{alert} function, we developed our crawler in such a way that allows it to interact with these pages but not get trapped by them.

We deployed the \tool Crawler on three different sites (our campus, Amazon's Elastic Compute Cloud~\cite{aws}, and on Linode's cloud~\cite{linode}). We provided each instance with the same set of 10,000 possible typosquatting domains, which we obtained by applying the typosquatting models of Wang et al.~\cite{Wang:2006:STD:1251296.1251301} on the top 200 websites according to Alexa, and a set of 3,000 shortened URLs belonging to ten popular ad-based URL shorteners. The crawler instances initiate the crawling process at the same time each day and collect and store all the aforementioned data. Note that \tool was originally relying just on domain parking in order to find technical support scams and we incorporated ad-based URL shorteners later in our study. We denote the exact date while analyzing the data in Section~\ref{sec:data-analysis}.

\vspace{0.5ex}
\noindent\textbf{Detector.} 
The Detector Module identifies the pages that are the most likely to be technical support scams based on a set of heuristic rules. We examined several heuristics, such as, having a redirection chain, showing consecutive \texttt{alert} dialogues, the presence of a phone number, and the presence of special keywords. After observing approximately a week's worth of collected data, we designed our heuristic which minimized false negatives and false positives as follows: If a page has any kind of popup dialogue, we check its content using an empirically constructed decision tree and based on the presence of carefully chosen sets of keywords, we score the page and mark it as malicious if the score is higher than a tuned threshold. 

To gauge the accuracy of our heuristic, we use random sampling to select three days (from the 250 days that are crawlers are active) and manually analyze all page screenshots collected by \tool (17K screenshots), during those days. Through this process, we verified that our heuristic was able to capture \emph{all} technical support scams collected by \tool. Interestingly, we identified some scam pages that would use HTML to draw fake alert boxes when a user visited them. Our heuristic, however, can still detect them as they switch back to the native JavaScript alerts when the user attempts to navigate away from the page (aiming to trap the user on the same page). This manual inspection makes us confident that our heuristic can account for most, if not all, of the technical support scams that \tool was exposed to during the monitored period. 

\vspace{1ex}
\noindent\textbf{Liveness Checker.} The Liveness Checker is the final component of \tool which is responsible for tracking the lifetime of a scam page after it first appears in the crawler's feed. Every URL that the Liveness Checker receives from the Detector component, is added in a database of URLs that will be crawled on a daily basis. In addition, for every URL received, the Liveness Checker computes neighboring URLs that could be hosting a technical support scam page, e.g., removing GET parameters from a URL and iteratively reducing the resource-path until we reach the main page of a domain.
On any given day, a scam is considered to be ``alive'' if any of the above URLs responds with a page that matches our aforementioned scam-page heuristic. The lifetime of any given scam domain is the longest time period, in terms of days, that began and ended with a page marked as a technical support scam. We chose this definition to account for transient errors (support scam goes offline for one day) and for malvertising variance (same domain can first show a support scam, then a survey scam~\cite{clark2013there}, and then again a support scam).

\section{Data Analysis}
\label{sec:data-analysis}
In this section, we report on the data collected by \tool during a 36-week period, starting from September 1, 2015. \tool attempted to resolve 8.4 million domains and collected a total of 15TB worth of crawling data.

\begin{figure}[t]
\begin{center}
    \includegraphics[scale=0.43]{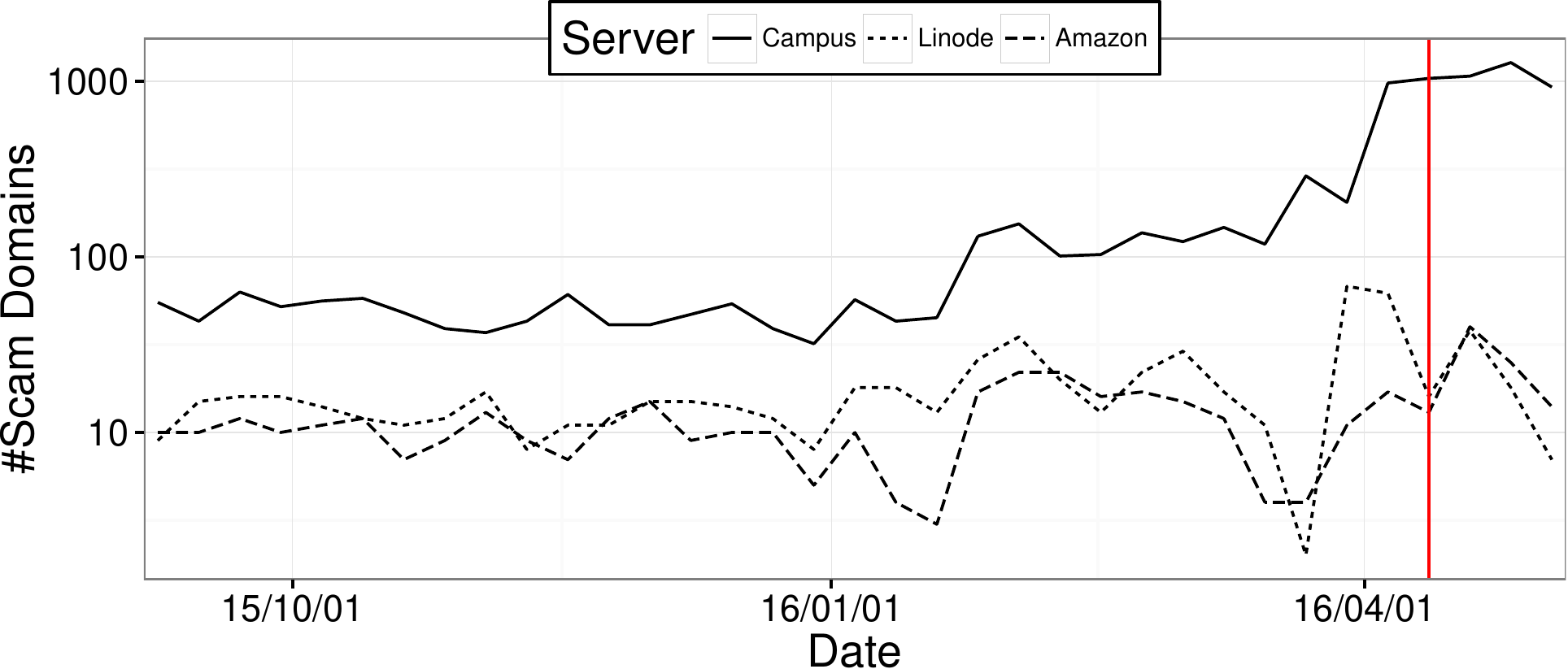}\\
    \includegraphics[scale=0.43]{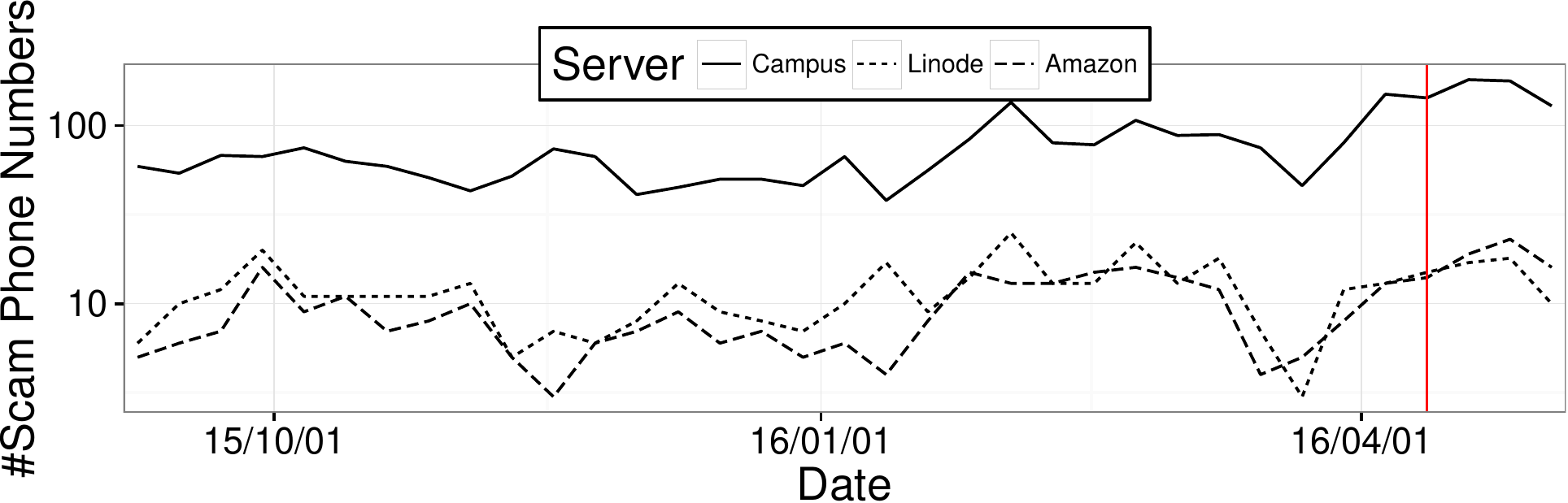}
	\vspace{-1ex}
    \caption{Number of unique weekly technical support scam domains (top) and phone numbers(bottom) recorded by each \tool instance during our 36-week monitored period. The vertical line denotes the week on which we adopted an extra source of malvertising pages, namely, that of shortened URLs.}
    \label{fig:scam_domains_day}
\vspace{-4ex}
\end{center}
\end{figure}

\subsection{Discovered scams}
Of the 5 million domains resolved by \tool, 22K URLs were detected as technical support scam pages, belonging to 8,698 unique domains. Figure~\ref{fig:scam_domains_day} (top) shows the weekly number of unique domains found by each of our three deployed \tool instances during our data-collecting period. One can see that, as time passes, technical support scams are becoming increasingly common reaching more than 1,000 unique domains per week in April and May 2016. Interestingly, phone numbers cannot keep up with that growth (Figure~\ref{fig:scam_domains_day}, bottom), suggesting that curbing the abuse of phone numbers will have a significant effect on technical support scams.

Another visible pattern is the great difference between the number of scam domains to which our campus-residing \tool was exposed, compared to the \tool instances located on Amazon's and Linode's hosting clouds. Since all crawlers were asked to crawl the same domains and none of the three \tool instances experienced any downtime during our monitored period, the only reasonable explanation is that the dubious advertising networks responsible for redirecting a user from a typosquatting page to a technical support scam page are using the user's IP address as a way of straightforwardly evading crawlers located on popular commercial clouds.

\begin{figure}[t]
\begin{center}
    \includegraphics[scale=.7]{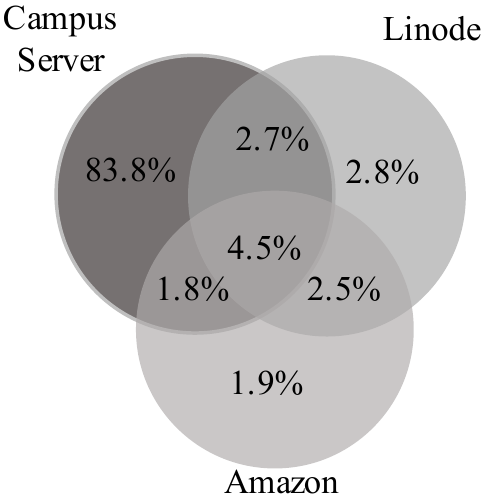}
    \caption{Venn diagram of Unique Phone Numbers Detected by \tool instances}
	\vspace{-6ex}
    \label{fig:phone-venn}
\end{center}
\end{figure}

An alternative way of looking at the unique scam domains discovered, is to consider the individual coverage of each of our three \tool instances. In terms of domain names, our campus-residing \tool, discovered 95.7\% of the domain names discovered by all three instances, with the Linode- and Amazon-residing \tool instances, contributing only 7.6\% of the overall unique domains. Similarly, the same campus-residing \tool instance, by itself, discovered 92.8\% of the total number of unique telephone numbers (see Figure~\ref{fig:phone-venn}). Overall, our results indicate that, because ad networks and attackers are location-aware, proxy-less servers located on popular commercial clouds, have only a small contribution in the discovery of scam pages and phone numbers.

Figure~\ref{fig:scam_domains_day} (bottom) shows the number of unique telephone numbers discovered each day and exhibits a similar behavior as Figure~\ref{fig:scam_domains_day} (top). Comparing the two figures together, one can see that while telephone numbers and domains are clearly correlated, the relationship between the two is not a 1-to-1 relationship. 
The reason for this is that scams located on different domains can be showing the same phone number, as well as the phone number on any given page can change between page loads. By inspecting some of the JavaScript code located in such pages, we found evidence of ``on-the-fly'', phone-number delivery.

By analyzing the code we found that scammers are abusing a pay-per-call management framework called Callpixel (rebranded as Retreaver). In Pay-per-call marketing, unlike pay-per-click (PPC) advertising, an advertiser will be charged if the ad visitor makes a call to a tracked phone number. Retreaver, as a pay-per-call management framework, can track and tag visitors based on various features, such as, browser profile, and geolocation, and provide them with customized toll-free numbers assigned ``on-the-fly'' from a pool of numbers. As the number is only available to a specific campaign at a specific time, Retreaver can tag the incoming calls, forward them to the call center configured for that campaign, and log it to calculate the conversion rate. 10\% of the domains in our corpus used variations of JavaScript code to use the Retreaver API. While this novel marketing model is designed for legitimate businesses for delivering highly targeted pay-per-call campaigns, it is abused by the scammers to generate fresh numbers dynamically. Figure~\ref{fig:chrome-offsetwidth}, shows a sample scam snippet which reads browser properties (operating system, user agent, and language) and calls the Retreaver API to get a JSON file containing a telephone number.

Lastly, by analyzing the data collected by the Liveness Detector module of \tool, we discovered that the lifetime of scam domains forms a long-tail distribution. Specifically, 27\% of the domain names are reachable only for a single day after they are first discovered by \tool, while 43\% of the domains are reachable for up to three days. At the same time, 7\% of the discovered domains were reachable for more than 40 days indicating that these are successful in avoiding unwanted attention and take-downs.

\begin{figure}[t]
\begin{lstlisting}[basicstyle={\scriptsize\ttfamily},language=JavaScript,showstringspaces=false,tabsize=2,frame=tbrl]
var ran = false;  
function loadNumber() {  
	if (!ran) {
		//Default numbers in case script fails
		var default_number = "(877) 292-3084"; 
		var default_plain_number = "8772923084"; 

		//Initiates new instance of specific campaign
		var campaign = new Callpixels.Campaign({campaign_key: '43019bb72cd5ecc4e3b33902645dd4d6'}); 
		//Script collects information about the user and the affiliate ID of the scammer
		var tags = {};
		var source_host = 'https://gyazo.com/714[...]b7915';
		var affiliate_id = '1';
		var clickid = 'Rb10lsaOkY';
		var browser = 'Firefox';
		var browserversion = '25.0';
		var country = 'US';
		var os = 'Windows';
	  	[...]
		//Populates an object with the gathered information
		tags = {
			a: affiliate_id,
			clickid: clickid, 
			source_url: source_host,
			browser: browser,
			browserversion: browserversion,
			country: country,
			os: os,
			[...]
		};
		//Function that retrieves a dynamic number
		campaign.request_number(tags,
			function (matching_number) {
				//Stores the dynamic number in global variable
				number = matching_number.get('formatted_number');
				plain_number = matching_number.get('plain_number');
				window.callpixels_number = matching_number;
			}, 
			function (error) {
				number = default_number; 
				plain_number = default_plain_number;
			} 
	); 
	ran = true; 
	//Shows the new number to victim user
	var number = "1 "+number;
	FormattedNumber1.innerHTML = number;
	[...]
}
window.onfocus = loadNumber();
\end{lstlisting}
\vspace{-2ex}
\caption{Partial JavaScript code that shows the dynamic fetching of a toll-free number based on the current victim's attributes, and the fallback logic in case the dynamic fetching fails.}
\vspace{-6ex}
\label{fig:chrome-offsetwidth}
\end{figure}

\begin{figure}
\centering
\begin{minipage}{0.49\columnwidth}
\centering
\includegraphics[scale=0.09]{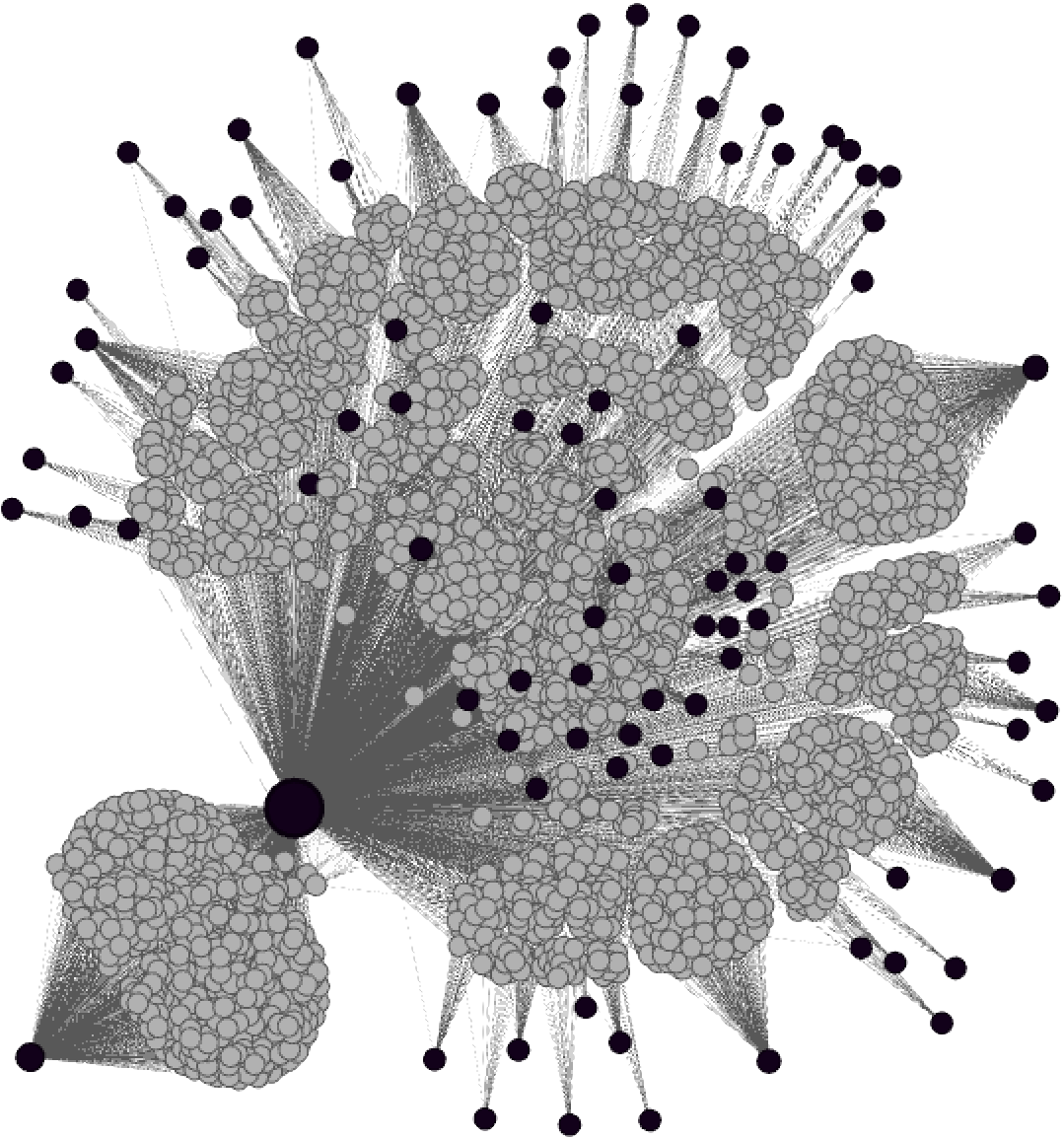}
\end{minipage}
\begin{minipage}{0.49\columnwidth}
\centering
\includegraphics[scale=0.1]{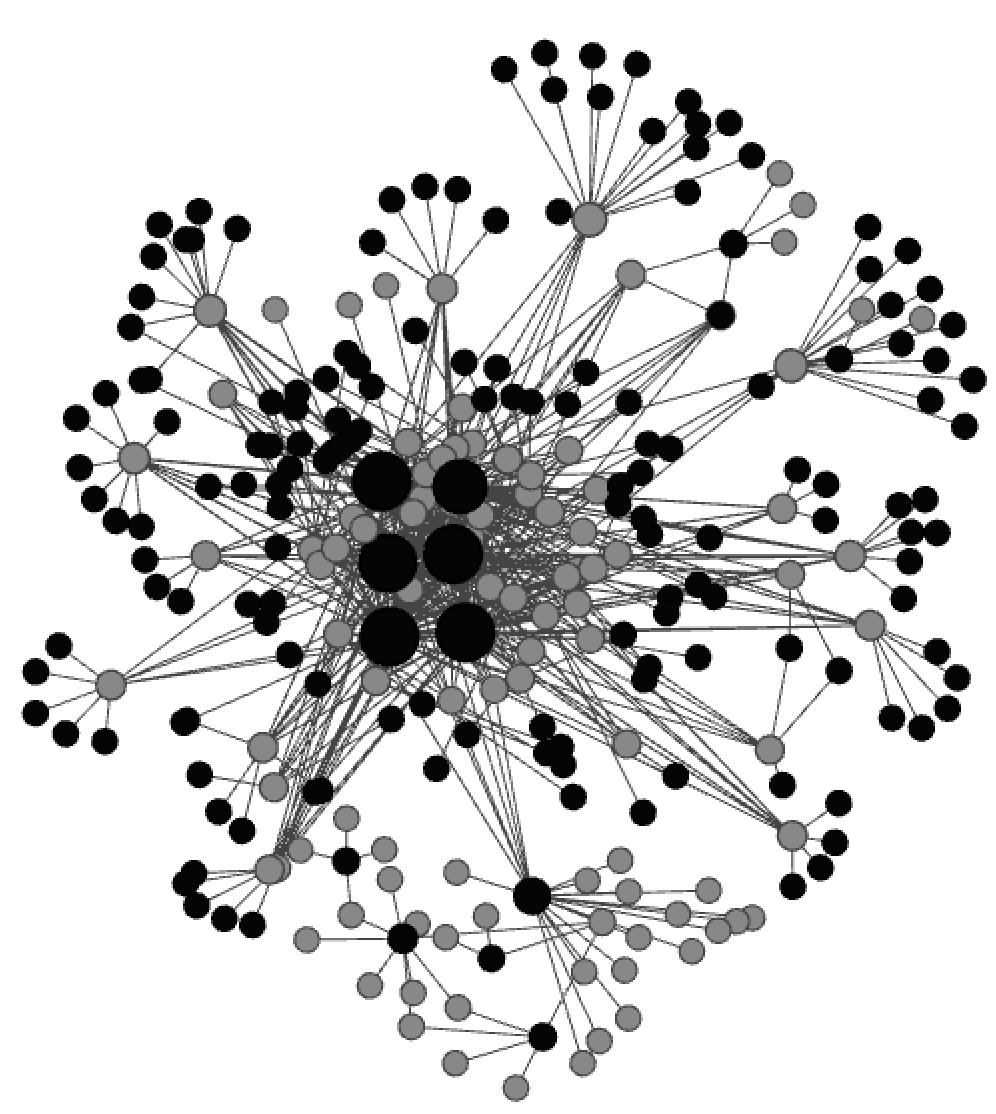}
\end{minipage}
\vspace{-2ex}
\caption{Two samples of technical support scam campaigns. The left graph shows the relationships between unique domains and phone numbers. The right graph shows the relationship between unique, TLD+1 domain names and phone numbers. Black and gray nodes represent phone numbers and domain names/TLD+1 domains respectively and size of a node is proportional to the node degree.}
\vspace{-3ex}
 \label{fig:net_graph}
\end{figure}

\begin{figure}[!th]
\centering
\includegraphics[scale=0.37]{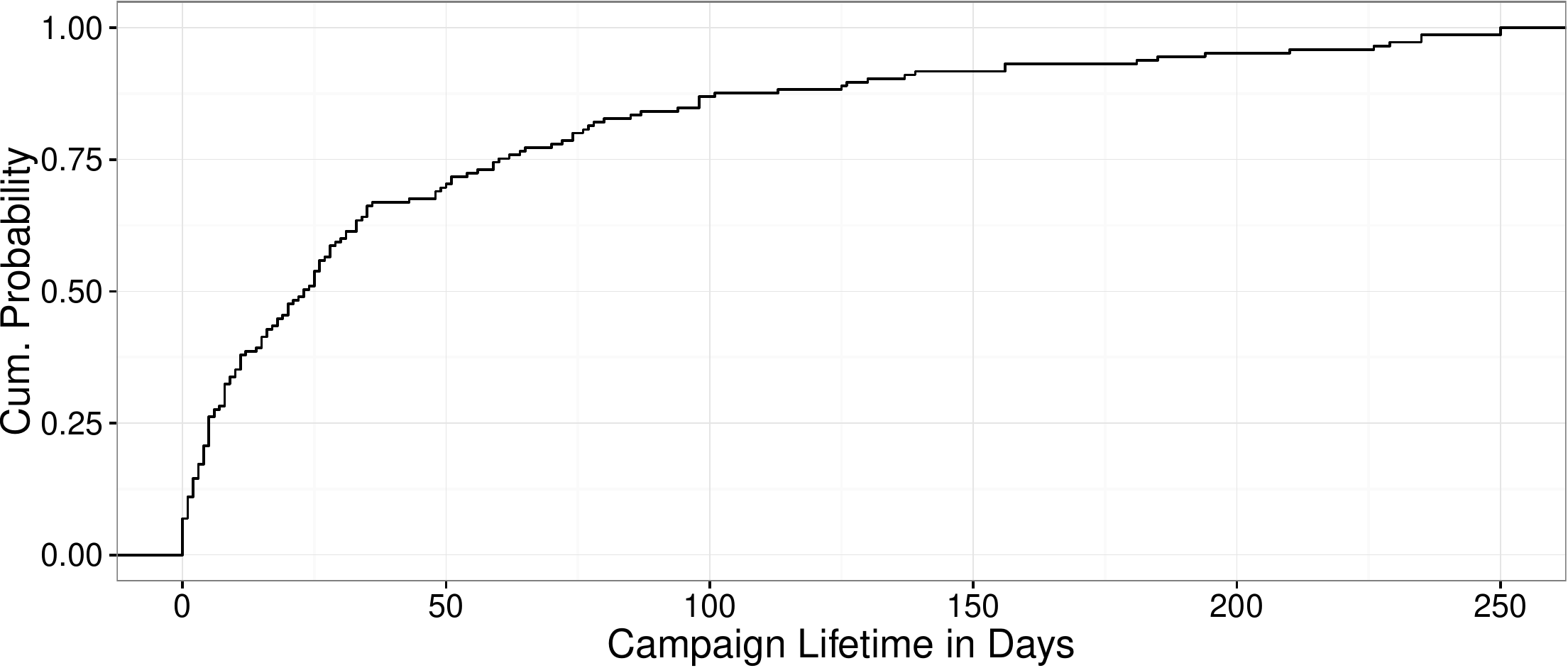}
\vspace{-1ex}
\caption{CDF of the lifetime of scam campaigns.}
\vspace{-4ex}
 \label{fig:campaign-liveness}
\end{figure}

\subsection{Domain names}

Of the total 8,698 unique scam domains collected by \tool, 17\% are completely human readable making extensive use of words that either imitate a legitimate brand, or attempt to scare the victim. The five most frequently used words in domains were:  \texttt{techsupport}, \texttt{alert}, \texttt{pc}, \texttt{security}, and \texttt{windows}. 83\% of the domains contained at least one random string and 6\% belonged to Content Delivery Networks (CDN) such as CDN77, CDNsun, KeyCDN, and MetaCDN.

Although the primary goal of CDNs is to provide high availability of static content, scammers abuse them as a way of obtaining free or near-free hosting for their scam pages. Content Delivery Networks, such as, CDN77, CDNsun, and KeyCDN offer free services without requiring a phone number or a credit card. In addition, every uploaded scam page gets its own random-string-including URL which can not be guessed and thus cannot be preemptively blacklisted (blacklisting the entire CDN-controlled domain would cause collateral damage).

Technical support scam domains are unusually long. A t-test on the distribution of domain length of 8K scam domains (with an average length of 76$\pm$56) and the top 8K Alexa domains (with an average length of 12$\pm$3.5) results in a very small p-value (p$<$0.05) which indicates that the difference is significant. By inspecting a sample of the scam pages hosted on long scam domains, we found that scammers make use of long domains to, among others, evade the built-in mechanism of the browsers for suppressing pop-ups. We discuss these techniques further in Section~\ref{sec:page_contents}.

The set of collected scam domains, after removing CDN entries, maps to 1,524 TLD+1 domains resolving to 685 unique IP addresses. This reduction in the size of hosting providers, confirms the use of shared-hosting as a way of getting cheap domains and hosting which can be easily changed to evade blacklisting. The majority of scam-page hosting is done in the US (88\%), followed by a long tail of various countries, such as, India and Netherlands. While India is not an obvious hosting choice, we show that many scammers seem to be operating out of it (Section~\ref{sec:calls}). We also mapped the IP addresses to AS names and found that 18\% of the scam hosts are using Cloudflare to hide their hosting server.

%\todo{Alex, whois clustering}
Lastly, we focused on the WHOIS records of scam domains. From the 1,524 TLD+1 domain names, we were able to extract 1,055 (69.2\%) email addresses. The difficulty of parsing WHOIS data~\cite{Liu:2015:CLP:2815675.2815693} was compounded by the fact that 344 domains did not even have an email address listed in their WHOIS records. 589 email addresses belonged to multiple WHOIS privacy companies which stopped us from identifying common individuals behind different domains. By inspecting the remaining 466 addresses we noticed patterns of similar names, such as, \texttt{amitabb8@gmx.com}, \texttt{amitabb9@gmx.com}, \texttt{amitapp1@gmx.com} and \texttt{amitabb6@gmail.com}. To automatically cluster these emails we used the Levenshtein distance metric and grouped together addresses with a distance of less than 5. This resulted in the formation of 192 clusters, including 65 clusters with at least two email addresses and 6 with more than ten. The two largest clusters contained 60 (united by \texttt{supernetws[0-9]+@yahoo.com}) and 27 (united by \texttt{charmssprince@gmail.com}) domains respectively. Our results highlight that even though scammers attempt to hide from analysis systems, a large-enough corpus of scam domains may still allow the grouping of seemingly unrelated scams.

\subsection{Phone numbers and their relationship to domains}

Since phone numbers are a crucial part of technical support scams, we used a public database of toll-free numbers~\cite{tollfreenumbers} to get more information about them. There, we discovered that even though the 1,581 toll-free numbers belong to 15 different telecommunication providers, more than 90\% belong to only four providers (Twilio, WilTel, RingRevenue, and Bandwidth) which indicates that scammers are abusing some providers significantly more than others. Moreover, we discovered 77.5\% of the phone numbers were activated less than one year ago and none of the vanity terms associated with the collected numbers is related to tech support.

To gain insights on the N-N relationship between scam domains and phone numbers appearing on scam pages, we plotted their network graph. In this graph, an undirected edge between a domain name and a phone number exists, if the phone number was advertised by the domain name during the time period of our experiment. The resulting graph contains 582 connected components of various sizes, of which 216 connected components have more than 5 nodes. A sample of the connected components is depicted in Figure~\ref{fig:net_graph} (left). As one can notice, the same numbers are reused across a set of domain names and, vice-versa, a domain may advertise different phone numbers over its lifetime.

\begin{table}[t]
    \caption{Characteristics of the top five campaigns. D: Domains, P: Phone numbers}
    \label{tbl:tld_prefix}
    \centering
    \begin{adjustbox}{max width=\columnwidth}
    \begin{tabular}{ccp{1.2cm}p{0.9cm}p{.8cm}p{0.8cm}p{1.2cm}p{1cm}}
        \hline
        \shortstack[c]{\textbf{\#D}} & \textbf{\#P} & \textbf{TLDs} & \textbf{Prefixes}& \textbf{\#IPs/\#ASs}& \textbf{Country}& \textbf{Top AS or CDN}&  \textbf{Lifetime (days)}  \\
         \hline
        3714 & 93 & net, com & 855, 844, 888, 877 & 35/3 & US, NL & CloudFlare &  64\\
        \hline
        513 & 96 & biz, net, com, in, us, xyz, space, website, info, club, online, me, cf, ga, org, co, tk, ca, site & 844, 877, 855, 866, 888, 800 &93/20& US, IN, FR & Cloudflare, GoDaddy & 250\\
        \hline
        173 & 359 & space, info, com, org, net & 888, 855, 844, 877 & 42/7 &  DE, FR, US & cdn77, cdnsun, metacdn, keycdn & 235\\
        \hline
        145 & 164 & info, help, online, website, com, xyz, in, net & 888, 844, 877, 855, 800, 866 & 33/9 & US, IN, NL& Amazon & 185\\
        \hline
        68 & 15 & net, com, org, info & 844, 888 &1/1&US& 1 and 1 & 250\\
        \hline
    \end{tabular}
    \end{adjustbox}
\vspace{-4ex}
\end{table}

To identify connected components which are more representative of scam campaigns, we merge the domain nodes that have the same TLD+1 domain and replot the network graph. The new graph contains 434 connected components while the phone nodes and domain nodes have an average degree of 2.8 and 2.5 respectively. The maximum degree of phone nodes is 173, and the maximum degree of domain nodes is 34. One sample of a connected component in this graph which represents a technical support scam campaign is plotted in Figure~\ref{fig:net_graph} (right). One interesting characteristic of this subgraph is that the center six phone numbers are connected to almost all of the campaign's domain names. After investigating these specific scam pages, we discovered that these numbers are the default numbers that would be used by the scam page in case an error happens during the on-the-fly retrieval of a new phone number.

We estimate the life time of scam campaigns by adding timestamps to the nodes of the network graph. We define the lifetime of a campaign as the difference between the timestamps of the first and last domain or phone number joined to the subgraph of the campaign. As Figure~\ref{fig:campaign-liveness} shows, the distribution of campaigns' lifetime is not normal and 69\% of the campaigns have a lifetime of less than 50 days. Even though the average lifetime is 45 days, there are campaigns with a life time of more than 250 days (the whole duration of our experiment). Moreover, assuming that the size of a campaign is equal to the size of its graph, there is a positive correlation (r=0.5) between the lifetime of a campaign and its size. We can, therefore, conclude, that larger technical support campaigns tend to be active for a longer time. 

Table~\ref{tbl:tld_prefix} shows the characteristics of the five largest campaigns and their estimated lifetime. The utilized toll-free prefixes, TLDs and hosting infrastructure differs among campaigns with the two first campaigns, besides having rich and diverse infrastructures, hiding their hosting servers behind Cloudflare. One can also see that many of these campaigns use cheap TLDs, such as, \texttt{.xyz}, \texttt{.space} and \texttt{.club}, to generate many variations of scam domains.

\begin{figure}[t]
\begin{center}
    \includegraphics[scale=0.25]{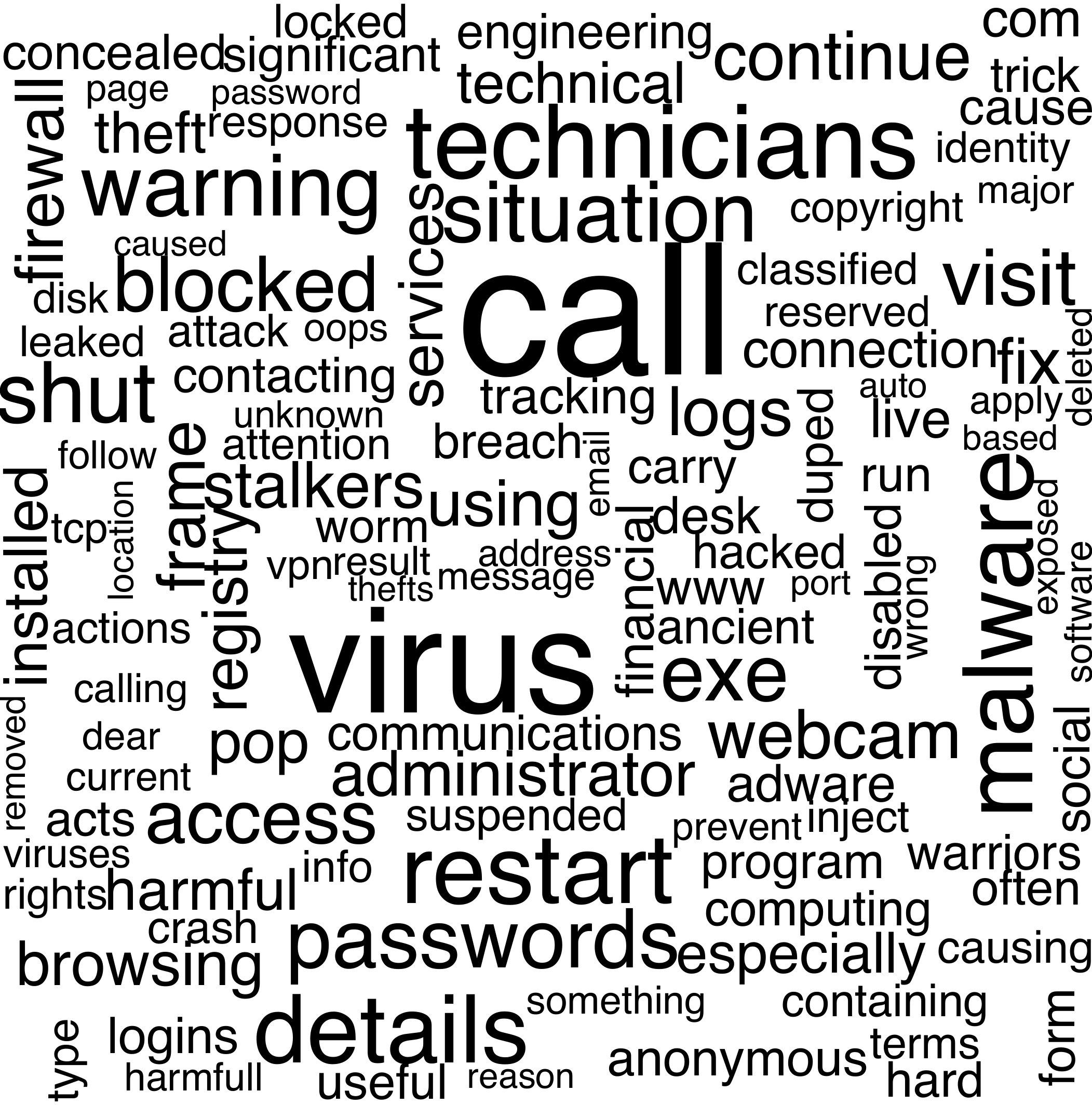}
%    \vspace{1ex}
    \caption{Word cloud based on the text contents of the gathered technical support scam pages}
	\vspace{-7ex}
    \label{fig:wordcloud}
\end{center}
\end{figure}

\subsection{Page contents} \label{sec:page_contents}

Scammers use specific words in the content of a scam page to convince the users that their machines are infected with a virus. Figure~\ref{fig:wordcloud} shows the most frequent words used in the scam pages in the form of a word cloud, where the size of each word is correlated with the number of times it appeared in our collected corpus of technical support scam pages.

Next to specific words, scammers also abuse browser APIs to increase the effectiveness of their scams. In Section~\ref{sec:background}, we discussed how scammers abuse \texttt{alert} dialogues to make it hard for users to navigate away. Some browsers, however, give users the ability to suppress \texttt{alert} dialogues, if a page is abusing them. For instance, in Google Chrome, if a page uses two back-to-back \texttt{alert} dialogues, the browser adds to the second \texttt{alert} dialogue, a checkbox that the user can check to ``Prevent this page from creating additional dialogs." 49\% of the collected scams were using very long \texttt{alert} messages, padded with whitespaces and new lines in an attempt to elongate the \texttt{alert} dialogue to a point that the newly added checkbox would be out of the user's view. The rest were trying to bypass the \texttt{alert}-dialogue threshold, by using multiple event handlers, launching \texttt{alert} dialogues from each one, in combination with the creation of new pop-up windows and subdomains. It is also worthwhile to note that Internet Explorer does not offer such a mechanism and thus a malicious webpage can keep on launching \texttt{alert} dialogues without the user being able to stop them, or navigate away while a dialogue is shown.

Lastly, we observed that 87\% of the discovered scam pages were using HTML audio tags, to automatically launch repeating audio clips that either sounded like an alarm, or were text-to-voice tracks, highlighting the severity of the problem and asking the user to call the listed technical support number.

\subsection{Sufficiency of existing blacklists}
\label{sec:blacklists}
%\todo{Alex fill this in, both domains as well as phone numbers}

Technical support scams require both domain names as well as (toll-free) phone numbers. As such, one could reason that, in contrast with most other attacks where only malicious domains are utilized, defenders have two chances to protect users via blacklisting: one by blacklisting domains, and one more by blacklisting phone numbers. We evaluate popular blacklists and show that existing blacklisting efforts fall severely short of capturing scam domains and phone numbers.

\begin{figure}[t]
\centering
\includegraphics[scale=0.15]{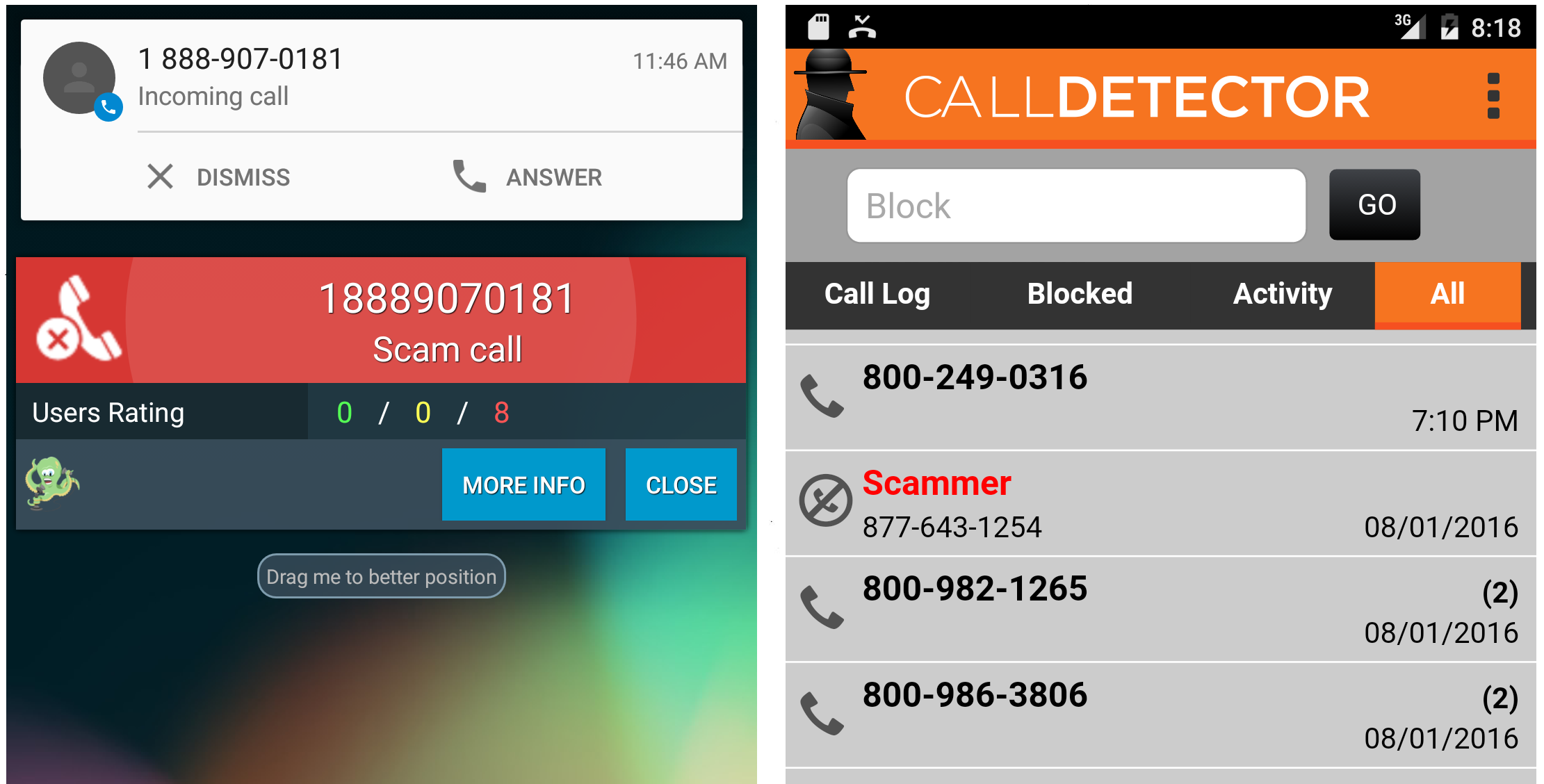}
\caption{Screenshots of phone lookup mobile apps when receiving calls from scam numbers: Should I Answer? (left), CallDetector (right)}
\vspace{-5ex}
\label{fig:apps}
\end{figure}

\noindent\textbf{Domain blacklisting.} First, we check our collected 1,524 TLD+1 scam domains against a combined set of popular blacklists including hpHosts~\cite{hphosts},  
suspicious domains by SANS~\cite{sans-malware}, malwaredomains~\cite{malwaredomains}, malwaredomainlist~\cite{malwaredomainlist}, Malc0de database~\cite{malcode-list} and, abuse.ch~\cite{abusech}. We use snapshots of these databases starting from 2014, which overall contain records for approximately 370K domains and IP addresses. Surprisingly, out of 1,524 scam domains, only 108 (7\%) were blacklisted. Moreover, out of the 108 blacklisted domains, only 16 were already blacklisted on the day that \tool first detected them. The rest were blacklisted, on average, 38 days \emph{after} \tool's detection. We also resolved the scam domains and checked whether their IP addresses were blacklisted. From the 685 resolved IP addresses, only 28 (4\%) were already present in one of the aforementioned blacklists.

Second, we repeat our experiment using VirusTotal's domain tools. There, we discovered that 974 of 1524 TLD+1 domains, i.e., approx. 64\%, were detectable by, on average, 3.25 AV engines on VirusTotal. Since VirusTotal does not show the date of first discovery of a malicious domain, we cannot calculate the exact fraction of domains which \tool discovered before AVs. Moreover, since VirusTotal houses 68 different AV engines, we argue that the vast majority of AV users are likely not going to be protected against technical support scams, even if they are accessing one of the 974 domains that were detectable by VirusTotal.

\noindent\textbf{Phone blacklisting.} There are different kinds of abuse possible via phone calls, including spam, scam, and extortion. Consequently, many phone-lookup services exist that keep databases of malicious phone numbers. These databases are typically crowdsourced relying on users to submit complaints. In recent years, in addition to websites that users can utilize to look up phone numbers, there exist mobile apps that give users real-time information about a number that is calling them.

To assess whether these databases include phone numbers involved in technical support scams, we scraped six websites searching for complaints for any of our \tool-detected 1,581 phone numbers. Moreover, we investigated how well the five most popular caller-id and fraud-protection Android apps were able to detect technical support scam numbers. 

To avoid reverse engineering each app, we opted to install each app in an Android emulator and simulate calls originating from each of the 1,581 scam-operated phone numbers. Some apps provided a call log which showed whether any given caller was a scammer (Figure~\ref{fig:apps} right). For these apps we simulated batches of calls and inspected the number of detected scams after each batch. For the apps that only showed a warning when a user was receiving a call (Figure~\ref{fig:apps} left), our tools captured a screenshot of the device during each simulated call and then used perceptual hashing to create clusters of similar screenshots. At the end of this process, we inspected the resulting clusters and manually labeled them.

\begin{table}[t]
    \caption{Presence of the collected 1,581 technical support scam phone numbers in popular phone lookup databases: website searches (white), mobile in-app alerts (gray)}
    \label{tbl:bphones}
    \centering
%    \begin{adjustbox}{max width=\columnwidth}
    %\begin{tabular}{p{2cm}cp{2cm}p{4cm}}
    \begin{tabular}{|l|c|l|}   
        \hline
        \textbf{Database} & \textbf{\% TSS numbers} & \textbf{Claimed Size}  \\
        \hline
        %\multicolumn{3}{|c|}{Website Searches}\\
        %\hline
        mrnumber.com & 19.9\% & 1.5 billion numbers\\
        800notes.com & 18.5\% & Unknown \\
        numberguru.com & 1.0\% & 29 million lookups\\
        \rowcolor{Gray}
        Should I Answer? & 0.5\% & 640 million lookups \\ 
        \rowcolor{Gray}
        Truecaller & 0.5\% & 2 billion numbers \\ 
        \rowcolor{Gray}
        Hiya & 0.3\% & 100 million numbers \\
        badnumbers.info & 0.2\% & 968,639 complaints\\
        \rowcolor{Gray}
        CallDetector & 0.1\% & 100,000 complaints monthly \\
        callersmart.com & 0.1\% & 5.9 million lookups\\
        \rowcolor{Gray} 
        Mr. Number & 0.1\% & 1.5 billion numbers\\ 
        scamnumbers.info & 0.1\% & 31,162 numbers\\ 
        \hline
        Together & 27.4\% & Unknown \\
        
        \hline
    \end{tabular}
%    \end{adjustbox}
\vspace{-4ex}
\end{table}

Table~\ref{tbl:bphones} presents the results. The best (in terms of coverage) phone lookup websites cover less than 20\% of technical support scam phone numbers, and for both \texttt{mrnumber.com} and \texttt{800notes.com}, more than 25\% of detected scam phone numbers were, in fact, detected by \tool on average 44 days earlier (comparing the date of discovery with the date of the earliest user complaint). Mobile apps perform significantly worse with all apps detecting less than 1\% of the 1,581 scammer-operated phone numbers. Mr.Number is an interesting case since the same company operates both the app as well as the website, yet the app warned us only for two of the technical support scam numbers. We suspect that this could be because of some overly strict procedure for verifying user-submitted complaints and converting them to warnings that users will see on their device. Even worse, some phone numbers were identified by mobile apps as legitimate ``computer'', ``windows'', or ``repair'' businesses (even ``Dell'' or ``McAfee''), ``technical support'', or ``facebook support'', and ``call centre'' with positive reviews. 

If users were to install all apps on their phones and inspect all websites before making a call, they would still cover less than 30\% of our collected scam numbers. By analyzing the text of complaints regarding known scam phone numbers on the evaluated websites, we found evidence that some scam numbers are also involved in other types of scams, including fake law firms, debt collectors, and IRS scams. We found 10 such cases on \texttt{mrnumber.com} and 24 on \texttt{800notes.com}. In terms of apps, Hiya app identified one of malicious numbers as an IRS scam.

\subsection{Estimating the number of victims, their location, and scammer profits}
%\todo{Najmeh fill this in}
Although we did not have access to the scammers' servers, we took advantage of the misconfiguration of Apache servers of some of the scam domains to collect data regarding their visitors. Specifically, we noticed that some hosting servers of technical support scam domains had enabled the \texttt{mod\_status} module of Apache servers without restricting access to it. This module provides an HTML page showing current server statistics including clients' requests, clients' IP addresses, server uptime, and total traffic~\cite{modstatus}. 

\tool checked each discovered domain for the presence of \texttt{mod\_status} module and appended those that were exposing it to a list of domains that were crawled \emph{every minute} by a separate crawler. In this way, by collecting more than 50GB of \texttt{mod\_status} data, we were able to monitor the activity of 142 scam domains over a period of two months. By analyzing the collected data, we discovered a total of 1,688,412 unique IP addresses visiting the monitored scam domains. 
On average, each scam domain received a total of 11,890 visitors (approximately 224 visitors per day). The maximum number of visitors of a scam domain was 138,514 and many popular domains had traffic originating from more than 68K unique IP addresses. 

By geolocating the IP addresses of the visiting users, we identify the following five most popular countries: United States (33.6\%), Australia (25.36\%), Singapore (22.40\%), Canada (7\%), and New Zealand (4.8\%). These results show that scammers are currently mostly targeting English-speaking countries. Note that this language-selection is a necessary property of technical support scams since scammers must be able to fluently speak the language of the victims who will contact them.   

Assuming that the conversion rate for victims landing on technical support scam pages (i.e. calling the listed phone number and buying the technical support scam package offered by the scammer) is the same as that of users paying for the ``full version'' of a fake antivirus (approximately 2\%, as calculated by Stone-Gross et. al~\cite{stone2013underground}), 33,768 of the 1,688,412 users have paid for unnecessary technical support. With the average price of a technical support scam package being \$290 (calculated in Section~\ref{sec:results}), just for the 142 \texttt{mod-status}-monitored domains, scammers have collected more than \$9.7 million from unsuspecting users.

\subsection{Summary of findings}
By building a tool (\tool) that can take advantage of malvertising in order to discover technical support scams, we were able to witness the growth and dynamics of these scams over an 8-month period. We found that scammers use thousands of domains and phone numbers to scare victims into calling them, and make use of blocking APIs and intrusive techniques to stop users from navigating away from their website. We showed that scams can be grouped into campaigns and discovered that the most successful scammers were able to work uninterrupted for the entire duration of our experiment. We identified the TLDs and telecommunications companies that are the most abused by scammers and found evidence of on-the-fly phone-number delivery. Finally, we witnessed the poor coverage of domain-name and phone-number blacklists and estimated that, just for a fraction of the monitored domains, scammers are likely to have made more than 9 million dollars by defrauding unsuspecting victims.

\section{Interacting with Scammers}
\label{sec:calls}
Even though the various measurements of the data collected by \tool (presented in Section~\ref{sec:data-analysis}) can be used to better understand the workings of technical support scam domains, they provide no insights on what happens when victim users, convinced that their machines are infected, call and interact with technical support scammers.

To shed light into this final but crucial part of technical support scams, in this section, we report on the data that we collected by posing as technically unsavvy users and calling 60 technical support scammers, while recording our entire interactions with them. During those interactions, we discover the way that scammers gain access to a user's machine, the methods and procedures that they use to convince the user of the purported infection, the average duration of each call, and the amount of money requested by each scammer.

\subsection{Experiment Preparation}

At its core, our study is an observational study. That is, we do not seek to apply different treatments to scammers and observe their effect. We merely seek to observe the methods that they use in order to defraud an average individual, with no security-related computer knowledge. Even though this defrauding happens on a daily basis, we unfortunately have no means of tapping into these conversations while they happen. For this reason, we had to pose as victims and record our interactions with the scammers. 

\vspace{1ex}
\noindent\textbf{IRB Approval.} Since scammers are human subjects, we applied to our institute's IRB and got permission to perform these recorded calls. Our approved application allows us to make use of deception (we are not revealing our true identities or intent to the scammers) and waive the requirement of consent (we do not ask the scammers whether they want to participate in our study). In addition, we convinced the IRB to allow us to avoid debriefing the scammers at the end of each call, to avoid information sharing from the side of the scammers that would place suspicion on future calls. Since scammers are already having these conversations with victims on a daily basis, our study does not incur any risk to their emotional, psychological, or physical wellbeing.

\vspace{1ex}
\noindent\textbf{Observed Environment.} The very first action that scammers perform after a victim user calls them, is convince the victim to give them remote access to their operating system. For our purposes, we made use of virtualization, where an installation of a Microsoft Windows 7 operating system was executing inside a Type-2 hypervisor. The use of virtualization not only allowed us to fully isolate a scammer's actions from critical infrastructure, but to also roll-back to a clean state of our operating system, after the end of each call.

To ensure that our VMs look like realistic user systems, we artificially aged our virtual machine, by installing different applications, downloading images and documents and placing them on the Windows desktop, and browsing many popular video sites, gaming sites, and news sites. We changed our system clock between different sets of actions so that some of our actions would appear to have occurred in the past, e.g., the timestamps of installed programs and files, and the dates available in our browsing history, placed these actions up to two years before the beginning of our experiment. Since we limited our visits and downloads to popular websites and applications, we are confident that our virtual environments were free from malware. Finally, we also removed obvious tell-tale signs of our virtualization environment by changing the appropriate Registry keys and the configuration of our virtual machine~\cite{virtualboxhide}. 
\subsection{Data Sources and Data Collection}

To select the phone numbers of technical support scammers, we randomly sampled the pages that \tool discovered and ensured that we did not call any number more than once. Note that \tool discovers technical support scam pages which claim that users are infected and flood the user with alert boxes, in an attempt to make the user unable to navigate away from the scam website. 
As such, we are confident that we never called a legitimate technical support number.

We used VoIP software with conversation-recording capabilities, packet-capturing software residing outside the VM for capturing the network traffic of our virtualized OS, and host-OS-residing screen recording software, for recording the visible actions of scammers, once they were given access to our VMs. After the collection of data from 60 different technical support scam calls, and the calculation of the statistics described in this section, we anonymized all copies of the collected data according to our IRB protocol.

\subsection{Script for our interactions}
Throughout our calls, we pretended to be average computer users who can use their PCs but have no computer knowledge beyond that. For example, we pretended not to know what an IP address is and, while we knew that having a virus is bad, we pretended not to know exactly what a virus does on our computer. We allowed the scammers to remotely connect to our system, following their instructions to the letter, and acted with shock, each time that a scammer would interpret something on our screens as the result of malware. Shortly after each scammer presented us with the pricing of his services, we either abruptly ended our calls, or found an excuse to politely hang-up. We never contradicted the scammers except during the last ten calls in order to discover how scammers react when users inform them that they are not convinced.

In a typical instantiation of this type of scam, once a victim calls the scammer and explains to him why she is calling, the scammer takes over the conversation. As such, we argue that even if each call is slightly different than the rest, the overall obtained results are aggregatable and generalizable to the population of technical support scam sessions. To quantify this phenomenon, we utilized a professional audio transcription service~\cite{rev.com} to obtain the text of five randomly selected calls. The average number of words spoken by scammers in each call is 1,367$\pm$407, whereas the average number of words from the victims (ourselves) is 530$\pm$172. In addition to the scammers speaking, on average, almost three times as much as the victims, the standard deviation also shows that regardless of the exact call, the variation of our answers was small compared to the variation of the scammers questions.

Lastly, we want to point out that we did not pay any scammer and therefore are unable to study scammers, \emph{after} they have charged users for unnecessary services. We chose not to pay scammers primarily for ethical reasons. As described later in this section, the average amount of money that a scammer requests is almost \$300. To get statistically significant numbers, we would have to pay at least 30 scammers and thus put approximately \$9,000 in the hands of cybercriminals, a fraction of which would, almost certainly, be used to fund new malvertising campaigns and attract new victims.

\vspace{-2ex}
\subsection{Results}
\label{sec:results}
\begin{table}[t]
    \caption{Remote Administration Tools used by scammers for getting access to their victims' machines}
    \label{tbl:rats}
    \centering
    \begin{adjustbox}{max width=\columnwidth}
    \begin{tabular}{|c|r|c|}
        \hline
        \shortstack[c]{\textbf{Remote Administration}\\\textbf{Tool}} & \shortstack[c]{\textbf{Websites}\\ \hspace{1ex}} & \textbf{Scammer abuse}\\
        \hline
        \multirow{3}{*}{LogMeIn Rescue} & \texttt{www.support.me}  & \multirow{3}{*}{60\%} \\
	& \texttt{www.lmi1.com} & \\
	& \texttt{www.logmein123.com} & \\ \hline
	CITRIX GoToAssist & \texttt{www.fastsupport.com}  & 21\% \\ \hline
	\multirow{2}{*}{TeamViewer} & \texttt{www.teamviewer.com} & \multirow{2}{*}{12\%} \\
	& \textit{Scammer-controlled} & \\ \hline
	\multirow{2}{*}{Other} & \texttt{www.anydesk.com} & \multirow{2}{*}{7\%} \\
        & \texttt{www.gethelp.us} & \\
        & \texttt{www.supremocontrol.com} & \\
        \hline
    \end{tabular}
    \end{adjustbox}
\vspace{-3ex}
\end{table}

\noindent\textbf{Remote administration tools}
Before a scammer can start convincing users that their machines are infected with malware, he must somehow get remote access to a user's machine. To that extent, the scammer must guide the user into downloading, installing, and allowing a remote administration tool which he will then use for his ``support'' session.

\begin{table}[t]
    \caption{Techniques used by support scammers in order to convince their victims of a malware infection}
    \label{tbl_techniques}
    \centering
\scalebox{0.9}{
    \begin{tabular}{|c|c|}
        \hline
        \textbf{Technique} & \textbf{\% Calls}\\
        \hline
        Stopped Services/Drivers & 67\\
        Event Viewer & 52\\
        Specific Virus Explained & 50\\
        System Information & 47\\
        Action Center &  40\\
        Fake CMD Scan & 40\\
        Netstat Scan & 40\\
        Installed/Running Programs & 35\\
        Browsing History/Settings & 27\\
        Downloaded Scanner & 17\\
        Reliability/Performance & 15\\
        Other (Temp, Registry) & 13\\
        \hline
    \end{tabular}
}
\vspace{-3ex}
\end{table}

Table~\ref{tbl:rats} shows the most popular tools abused by scammers for connecting to our machines. LogMeIn Rescue and CITRIX GoToAssist are web applications where a user visits one of the websites listed in Table~\ref{tbl:rats}, enters a code given by the scammer over the phone and downloads a binary that will eventually allow the attacker to remotely access a user's machine. TeamViewer and AnyDesk are stand-alone programs that a user must download and execute. Once the programs are running, both programs show a customer number and a PIN that a user must provide to the scammer in order for the scammer to connect to the user's machine.

In all cases, the scammers were abusing legitimate web applications and programs as part of their scams. Most of the aforementioned companies seem to be aware of this phenomenon and warn their users, typically through their websites, not to allow remote connections from people they do not trust. When these messages are pronounced, as in the case of TeamViewer, scammers incorporated these messages into their narratives in order to put us at ease. Other scammers, chose to self-host older versions of the programs that did not include these messages, thereby avoiding the warnings altogether.\\

\noindent\textbf{Utilized social-engineering techniques.}
The scammers used a variety of techniques to convince us of the purported infections and the need to purchase their support packages. Table~\ref{tbl_techniques} shows the most popular techniques used and the percentage of scammers that used each technique. We provide a brief explanation of the techniques that are not self-explanatory:

\noindent$\bullet$ \textbf{Stopped Services/Drivers.} 67\% of scammers loaded the list of Windows services and showed us that many services were stopped. While this is the normal state of a Windows OS installation, the scammers claimed that hackers have stopped these services and that is why they were able to get access to our machines.

\noindent$\bullet$ \textbf{Event Viewer.} Event Viewer is one of the administrative tools of Windows that shows general information about a system that could be used for troubleshooting purposes. The scammers treated the errors shown by this tool as a sign of hacker activity.

\noindent$\bullet$ \textbf{Virus details.} Some scammers, would conclude that our system is infected by specific malware, such as ``koobface'' or ``Zeus.'' They would then proceed to navigate our browser to pages explaining these threats and asked us to read out loud the section of each post describing the damage that the specific piece of malware does to its infected hosts.

\begin{figure}
\begin{center}
    \includegraphics[scale=0.58]{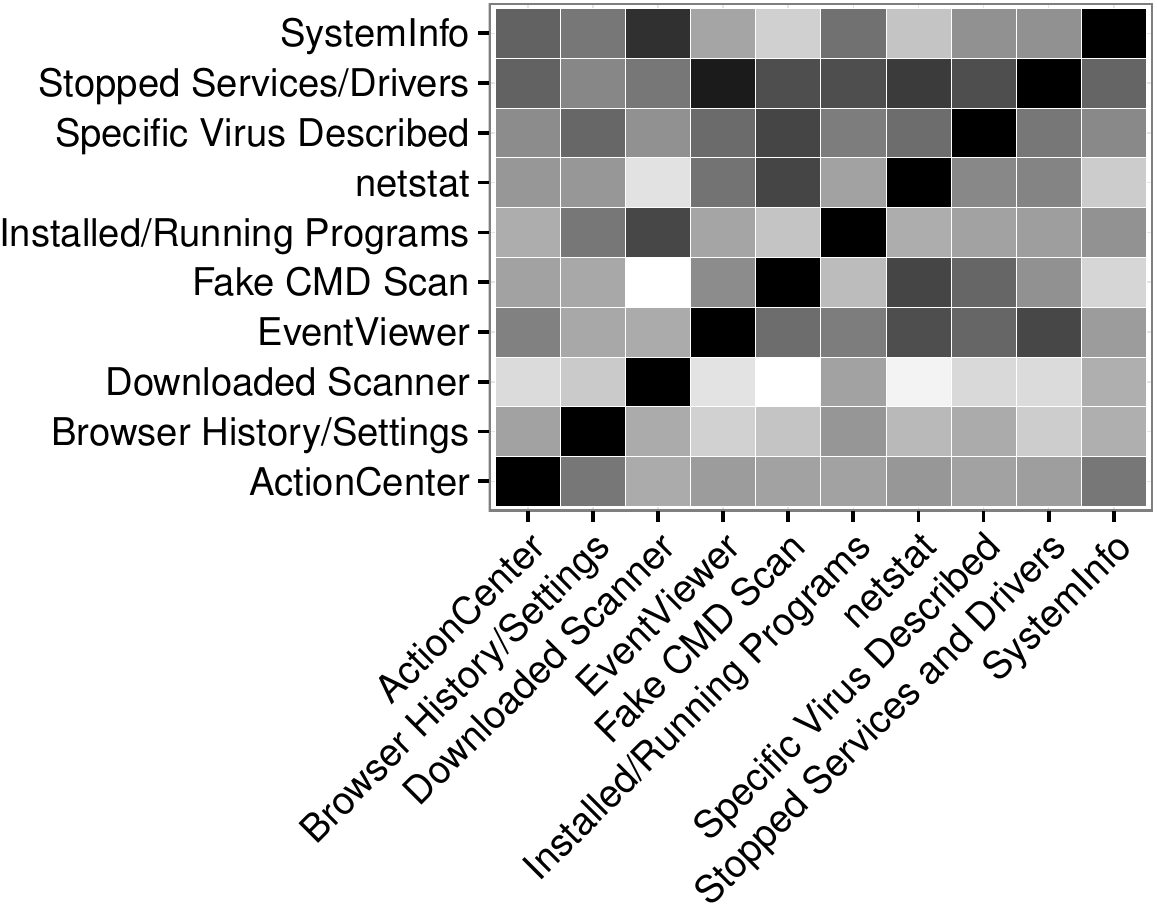}
    \caption{Heatmap showing the conditional probability (ranging from white to black) of the ten most often used social engineering methods. Note that because, in general, $P(A|B) \ne P(B|A)$ the heatmap is not symmetric along its diagonal.}
    \label{fig:techniques_heatmap}
\vspace{-6ex}
\end{center}
\end{figure}

\noindent$\bullet$ \textbf{Netstat scan.} 40\% of scammers utilized the \texttt{netstat} utility to convince us that our machine is already occupied by hackers. Specifically, they claimed that each non-local, TCP connection listed in the output of \texttt{netstat} was an attacker who had either already connected to our machine (entries with an \texttt{ESTABLISHED} status), or was currently trying to connect (entries with a \texttt{TIME\_WAIT} status).

\noindent$\bullet$ \textbf{Fake CMD scan.} One of the more creative techniques was the use of verbose command-line utilities as fake virus scanners. 40\% of the scammers utilized a command such as \texttt{``dir /s''} which lists files and folders present on a specific path of the filesystem. 
While the program is producing output, the scammer types or copy-pastes text in the command-line window, that will only appear \emph{after} the program is done executing. As such, at the end of the program's execution, the user suddenly sees text that claims that a virus has been discovered which he is likely to attribute to the ``scanning'' program that was just executing. This technique is likely one of the most convincing ones because i) it does not need interpretation (common messages used were ``Virus detected'' and ``System at Risk'') and ii) as far as the user is concerned, it is his own operating system that produces this message, rather than a downloaded third-party tool.

\noindent$\bullet$ \textbf{Performance.} Many scammers used system information tools to discover the type of CPU and amount of RAM available to our system. They then praised the hardware of our machine before proceeding to search for infections. This was typically done to convince us that spending money for the removal of malware was worth the cost since it would allow us to keep using our machine for many years before we would need to purchase a new one.
\vspace{1ex}

\noindent Overall, while we were able to identify techniques commonly used by scammers, we were impressed with the scammers' creativity in finding status messages that were already present on our system and attaching an infection meaning to them. Figure~\ref{fig:techniques_heatmap} shows how often scammers used two social engineering techniques together. There, we use the recorded frequencies to calculate their respective conditional probabilities.

\begin{figure}[t]
\begin{center}
    \includegraphics[scale=0.35]{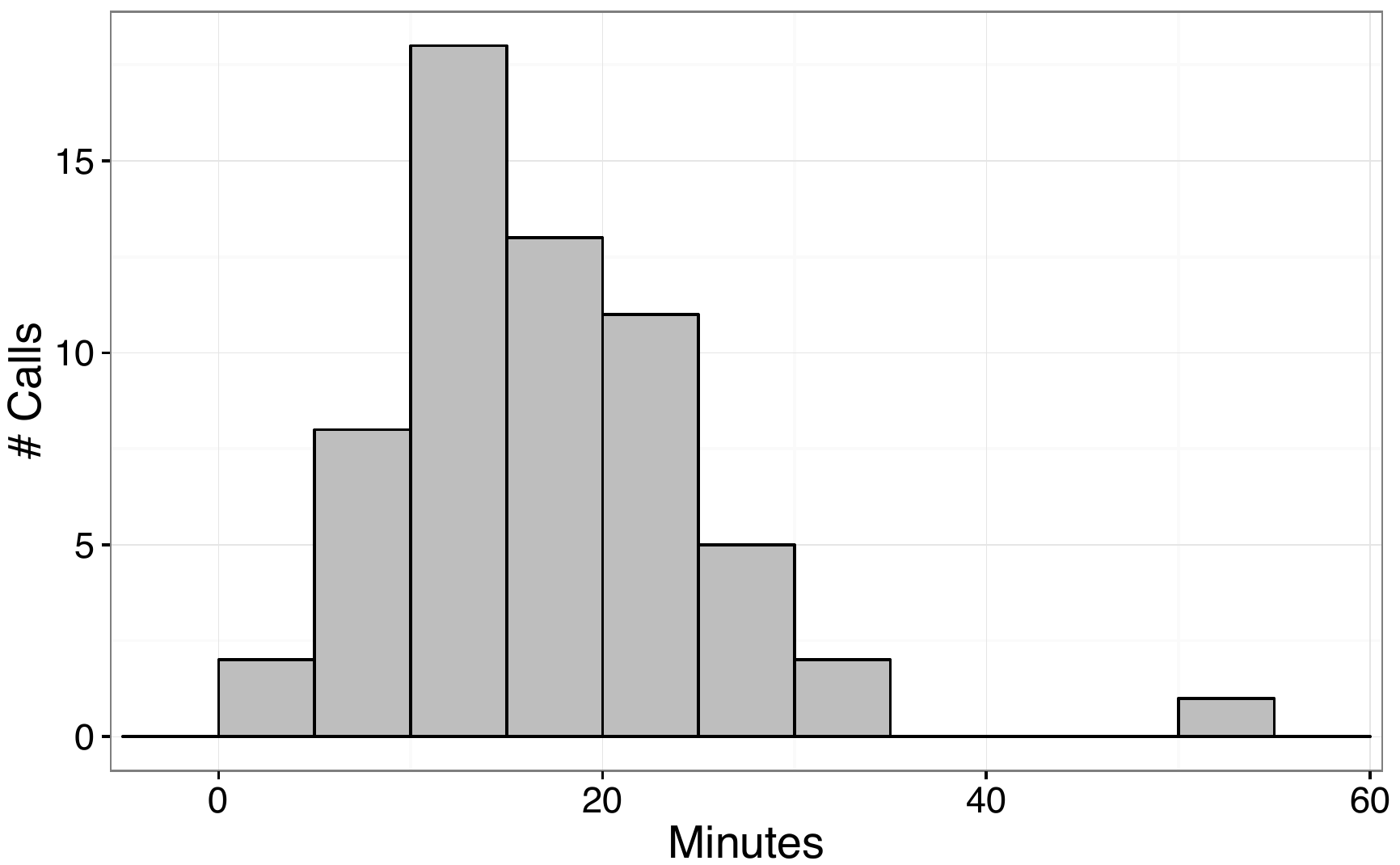}
    \caption{Distribution of the time duration between the beginning of a call, and the time when technical support scammers presented us with the pricing options for their services}
    \label{fig:time-distribution}
\vspace{-6ex}
\end{center}
\end{figure}

\vspace{0.5ex}
\noindent\textbf{Duration of calls.}
Figure~\ref{fig:time-distribution} shows the distribution of the time duration between the beginning of a technical support scam call, and the time when a scammer offered his services in exchange for money. The average duration of that interval is 17 minutes, and the distribution is approximately normal. In only a few cases, the scammers first told us the amount of money that they will be charging (around the second and third minute of our conversation) and \emph{then} proceeded to ``diagnose'' our machine.

Overall, one can see that the scammers are by no means in a hurry to convince users and defraud them. They take their time to slowly guide their victims into installing a remote administration tool, clicking through all the security dialogues, and giving them access to their machines. Once they have access, they slowly work their way through different Windows tools, showing their output to users and interpreting that output for them. It is likely that scammers know that the more time they take to convince a user about an infection, the more successful they will be when they ask for a compensation for their services. Figure~\ref{fig:time-distribution} also provides an indication of the amount of work necessary in order to obtain real-world data from technical support scam calls. Specifically, for the 60 calls recorded and analyzed, we spent, during a one-month period, more than 1,300 minutes (22 hours) just interacting with scammers, excluding dropped calls, out-of-order numbers, analysis of the recordings, and verification of our findings.

Since scammers control the vast majority of the conversation, we opine that the distribution of time shown in Figure~\ref{fig:time-distribution} will be generalizable to the population of victims. Therefore, this distribution can be of immediate value to telcos and the FTC. Specifically, given a list of numbers operated by scammers, telcos can straightforwardly produce metadata of their customer base that has called any one of the numbers of technical support scams. The FTC can then prioritize take-down action by focusing on the scammers with whom victims were interacting for more than 41 minutes, that is, the mean of our distribution plus three standard deviations. Since the duration distribution is approximately normal, the mean $\pm$ three standard deviations should capture approximately 99.7\% of all pre-charge calls. As such, anyone interacting for more than 41 minutes, is likely a defrauded victim.

\begin{figure}[t]
\begin{center}
    \includegraphics[scale=0.45]{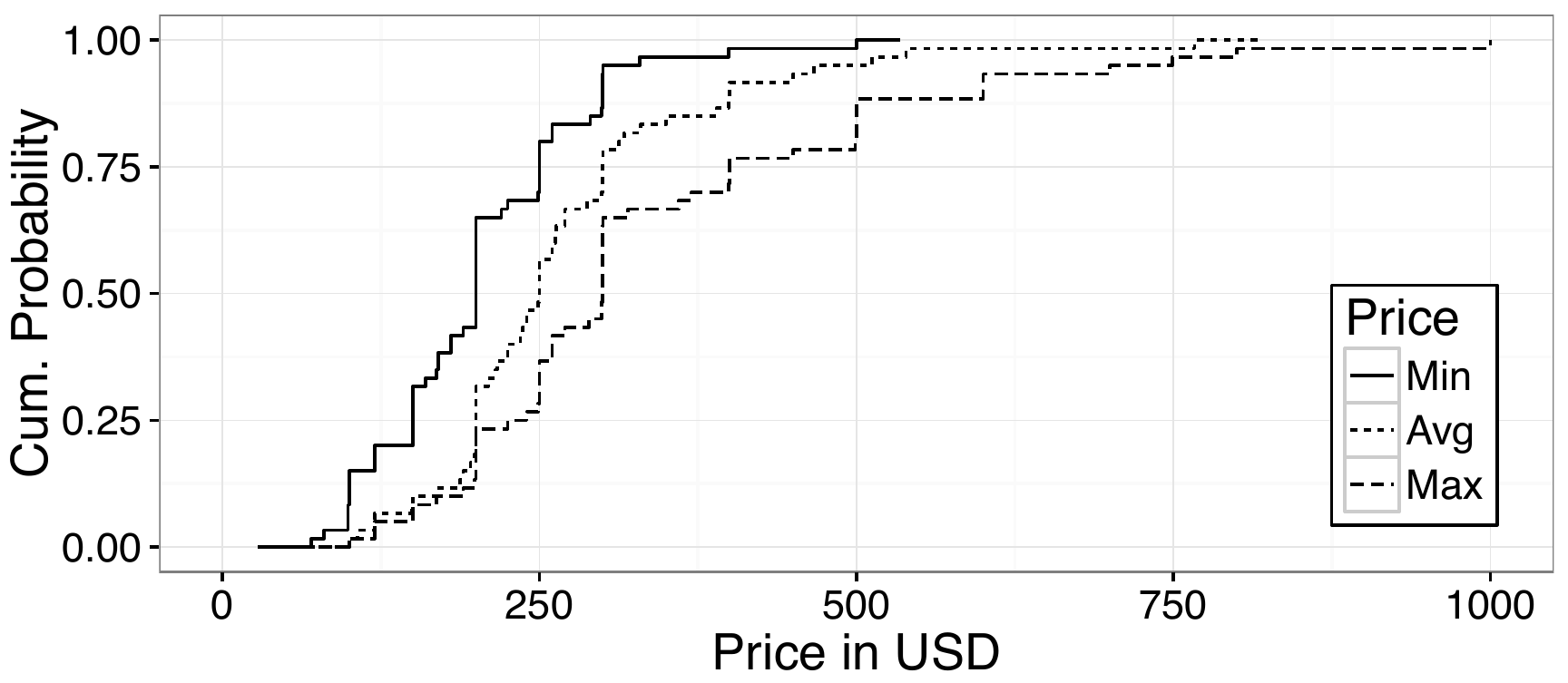}
    \caption{Requested charges for repairing our purportedly infected machines. Since most scammers offered us more than one support packages, we plot the ECDFs for minimum, average, and maximum amount requested.}
    \label{fig:prices-ecdf}
\vspace{-6ex}
\end{center}
\end{figure}

\vspace{0.5ex}
\noindent\textbf{Price of services.}
\noindent Once scammers felt confident that we were convinced that we are in need of their help, they then informed us about the price of their services. Most scammers offered us two to three different options with support packages ranging from a one-time fix, to multi-year support, ranging anywhere from \$69.99 to \$999.99. Figure~\ref{fig:prices-ecdf} shows the ECDF of the amount requested, split in its minimum, average, and maximum (average for any given scammer is the average price of all offered support packages). The average support price across all support packages and all scammers is \$290.9 with most scammers staying under \$500 for all of their support packages.

The prices of support packages were structured in a way where the middle one made the most financial sense. In fact, the times that we pretended to be willing to purchase their support and requested the cheapest option, the scammers would typically try to reason with us that the middle one was a better value-for-money offer. Interestingly, the price of services did not correlate with the time that scammers spend convincing us of our supposed infection (Pearson r=0.11).

\vspace{0.5ex}
\noindent\textbf{Freelance scammers vs organized call centers.}
At the outset of our study, we did not know whether technical support scammers are individual freelancers who supplement their income by single-handedly operating a technical support scam, or are part of an organized call center. Through the process of interacting with 60 different scammers, we are now convinced that most, if not all, scammers are part of organized call centers. 

Next to anecdotal evidence that we gathered during our interactions (e.g.  on one occasion, due to technical difficulties with our VoIP software, we called the same number three times in a row, and were greeted by a different person all three times), we conducted the following experiment: We replayed each recorded call and, instead of focusing on the scammer talking to us, we instead focused on background noises. While some scammers muted their microphones when they were not speaking, the majority did not. On 62\% of our calls, we were able to hear other people in the background, often recognizing phrases about security and malware that the scammer had just used in his own narrative. Therefore, our results indicate that the majority of scammers work in call centers, a fact which is corroborated by a recent interview of a technical support scammer on Reddit~\cite{tss-interview-reddit}.

\vspace{0.5ex}
\noindent\textbf{Estimating the size of call centers.}
Motivated by the finding that the majority of scammers operate out of call centers, we wanted to estimate the size of these call centers, i.e., how many scammers are ``hiding'' behind a single toll-free phone number.

To this end, we gathered 20 volunteers and explained to them the concept of technical support scams, the methods that scammers use, and the typical narratives of conversations with scammers. Each volunteer was given ten toll-free phone numbers operated by scammers (randomly selected by our pool of numbers) and a list of fake personae which they could assume when talking to the scammers (in our experience the majority of scammers request the caller's name and address before proceeding). The ten toll-free numbers were the same for all volunteers and they were instructed, guided by our signals and a projected stopwatch, to start calling each number at the same time. Each volunteer was instructed to engage with each toll-free number for a period of 90 seconds, either by talking to a scammer, or by waiting in a calling queue, or by redialing a busy number. Under the reasonable assumption that a scammer cannot be speaking to two people at the same time, this experiment essentially allowed us to estimate a lower bound of the size of a call center by counting the number of volunteers that were able to reach a scammer (either immediately or after waiting in a queue) in the measured 90-second period.

The average number of volunteers who were able to speak with a scammer across all ten studied phone numbers was 11, with the smallest call-center housing 5 scammers, and the largest one 19. Our results show that scammers can belong to various operations, ranging from small scale ones (call centers with 5 or 6 people) all the way to call centers that essentially occupied all of our volunteers (call centers with 18 or 19 people). As before, we argue that our method can be straightforwardly operationalized by the FTC and other law-enforcement agencies, for identifying the largest players in the technical support scam ecosystem, and focusing on them first.

\vspace{0.5ex}
\noindent\textbf{Scammer Location.}
Even though scammers access a user's machine via a remote administration tool that typically involves a centralized server relaying commands between the user and the scammer, it is possible that some tools still leak the scammer's real IP address to the user. To discover whether the remote administration tools utilized by scammers fit that description, we installed the tools on our machines and connected to them from another known IP address, while capturing the network traffic. We then analyzed the traces from our own connections and created packet signatures that reveal the connecting user's IP address.

Using this method, we recovered the IP address of 41 out of the 60 support scams. By geolocating these IP addresses, we discovered that 85.4\% of them were located in different regions of India, 9.7\% were located in the US, and 4.9\% were located in Costa Rica. While we cannot know with certainty that the scammers were not using VPNs located in these countries, we argue that they most likely are not since the recovered IP addresses do not belong to known VPN providers but rather to residential and corporate ISPs. In addition, the accent of the vast majority of the speakers with whom we interacted was Indian, matching our geolocation results. We reason that India is the most prevalent country, not only because of the relatively low average wage~\cite{indiasalary}, but also because India is already a popular choice for outsourcing call centers of English-speaking countries~\cite{callcenter1,callcenter2}. Consequently, we do not know whether the people running these call centers are the responsible ones, or are merely working for a third-party scammer who has outsourced the last part of the scams to them.

\vspace{0.5ex}
\noindent\textbf{Scammer Demeanor.}
In general, scammers exhibited a kind demeanor. They would patiently guide us through the steps for downloading their remote administration tool, giving us step-by-step instructions for the entire process. They would take no computer knowledge for granted, even to the point of explaining us that the Windows key is the one that ``looks like a flag'', between the Ctrl-key and the Alt-key on the bottom left of our keyboard. More than one scammer, after having explained to us that we are infected with malware, would open up Wikipedia pages trying to educate us of the meaning of words, such as, ``trojan'' and ``koobface.''%, and ``browser hijacker.''

To quantify how a scammer's behavior changes when faced with an expert user, in the last ten of our calls, after the scammers showed us ``signs'' of infection and offered their services in return for money, we contradicted them by explaining that we did not believe them. 60\% of the scammers remained calm and polite, and tried to convince us of the legitimacy of their company by showing us their websites and other online information. The remaining 40\% became rude and soon after that terminated the call, with one scammer setting a password to our virtualized OS before logging out.

\subsection{Summary of findings}

Through our interactions for over 22 hours with 60 scammers, we were able to precisely quantify many aspects of this last part of technical support scams. We discovered that scammers abuse popular remote administration tools (81\% of scammers rely on two specific software products), to gain access to user machines where they then patiently attempt to convince users that they are infected with malware. We found that, on average, a scammer takes 17 minutes, using multiple social engineering techniques mostly based on misrepresenting OS messages, to convince users of their infections and then proceeds to request an average of \$290.9 for repairing the ``infected'' machines. We explained why we are convinced that most scammers operate out of call centers, estimated the size of an average call center, and, using geolocation of the collected network traces, we found that scammers are likely to be operating out of some specific countries, more than others.

\section{Discussion and Future Work}

Given our findings in Sections~\ref{sec:data-analysis} and \ref{sec:calls}, 
we argue that technical support scams are a real and dangerous threat to the modern web. In contrast with other cybercrime methods, such as the stealing of credit card numbers and banking credentials, technical support scams do not need any additional monetization effort since, if the scam is effective, the victimized users will be \emph{willingly} accepting the charges and \emph{voluntarily} providing their private and financial information, over the phone, to scammers.

Even though systems that can automatically discover and detect these scams as soon as they arise, like \tool, are crucial, we opine that the threat of technical support scams can only be comprehensively subdued with the education of the public and additional help from browser vendors. In this section, we briefly describe these two areas of intervention and discuss the limitations of our work.

\vspace{0.5ex}
\noindent\textbf{User Education.}
User education has been a long-standing problem of security mechanisms and its lack has often been abused by attackers through social engineering. While certain problems, e.g., the expiration of an SSL certificate, or the problem of mixed inclusions, are admittedly hard to explain to a non-technical person, we argue that explaining the concept of technical support scams, is an easier endeavor. This is because, in technical support scams, there are no exceptions that the user must remember. A webpage cannot, by browser design, know that a user is infected and should never be using a flood of alerts with threatening messages to communicate with users. As such, educating the public that these pages should not be trusted is highly unlikely to cause harm to legitimate businesses, even the ones involved in remote technical support.

Public service announcements are already used by multiple countries as a way of raising awareness for health and safety issues, and would be an ideal vehicle for educating users about the dangers and characteristic signs of technical support scams. Even though the Internet Crime and Complaint Center called its warning of technical support scams a ``Public Service Announcement''~\cite{psa-ic3}, the announcement was only available via specific websites and thus far from the reach of the general population. At the same time, even though non-technical people can be educated to recognize technical support scams, we must also provide them with a simple way of navigating their browser to safety, away from webpages that abuse blocking, browser-provided APIs, such as the \texttt{alert} function, to keep users from navigating away.

\vspace{0.5ex}
\noindent\textbf{Browser Support.}
Given our reliance on the web, modern browsers try to provide high availability to users and a large degree of control to websites. In addition to blocking UIs, one specific feature that is, in general, desirable but has inadvertently become a tool in the hands of scammers is the remembering of open tabs in the case of a crash. Specifically, if we assume that, a non-technical user is trapped on a technical support scam page and, in a moment of desperation, reboots his machine, the browser will remember all the open tabs, including the one with the technical support scam, upon reboot. As such, the user will still be trapped and much more likely to call the scammers. Trying to outrun \texttt{alert} dialogues or killing the browser process and clearing recent history should not be something that we expect from everyday users.

To help users navigate to safety, we propose that browser vendors could all adopt one universal shortcut that users can utilize when they feel threatened by a webpage. Depending on the design, the browser can choose either to immediately close the current tab, or close all tabs and navigate the browser to a known safe page. The browser should ignore all event handlers and provide no way that a webpage could detect its unloading in time to launch a new window of the intruding webpage. Ideally, this shortcut combination would be communicated to the public through the aforementioned PSAs, allowing users to both recognize and defend against technical support scams. Lastly, we want to point out that such a shortcut could be useful beyond technical support scams, helping users quickly navigate away from websites that they find intrusive, such as shock sites, as well as helping them defend against any webpage that is trying to forcefully keep them from navigating away.

\vspace{0.5ex}
\noindent\textbf{Limitations.} We identify two limitations of our work: i) the potential evasion of \tool's detection heuristics, and ii) the non-completeness of our sources for discovering technical support scam domains.

First, we are well aware that, since we are operating in an adversarial environment, scammers may, as a result of our work, change their tactics to evade \tool's current detection heuristics. Note, however, that scammers do not have unlimited freedom in the techniques that they can use. Namely, if scammers want to trap a user's browser, they must identify blocking APIs (such as the currently abused \texttt{alert} method) and utilize those. Therefore, while an attacker could, for instance, use Flash to communicate their message (and therefore avoid matching our keyword-based heuristics) they would risk users just closing the offending tab. To account for language-based evasions, we are currently experimenting with a supervised machine-learning-based classifier, similar to the ones used for detecting spam. The premise is that the language that scammers use (consisting of warnings, errors, threats, and viruses) is different than the language of a typical webpage. We leave the discussion of this classifier and its accuracy for future work.

Second, we cannot unfortunately provide any guarantees of completeness of our malvertising-driven discovery approach. Such a guarantee would require information about the entire population of scammers and their traffic-delivering techniques, something that is likely infeasible. We chose to approach this problem pragmatically, by identifying malvertising as a key component of modern malware delivery and focusing on that. The presence of tens of different domain-parking and url-shortening services allows us to tap into multiple advertising networks which, in turn, exchange ads with even more smaller networks. We argue that since \tool is able to discover so much more than exists in current blacklists (as measured in Section~\ref{sec:blacklists}), we are, in fact, discovering a sizeable chunk of the true population of technical support scams.

%\vspace{-4ex}
\section{Related Work}

Our study was inspired by a series of blog posts and a whitepaper from an antimalware company which qualitatively analyzed technical support scams~\cite{malware-bytes-1,malware-bytes-2,mypchas}.  While these, and other blog posts have, in the past, analyzed a handful of scams, their studies are ad-hoc and their results are not generalizable. To our knowledge, no blog post has ever produced a repeatable methodology for finding scam pages in the wild, assessed the sufficiency (or lack thereof) of existing URL and phone-number blacklists, estimated the number of victims and the amount of money lost, or clustered phone numbers and their respective domains, all using a corpus of thousands of domain names and phone numbers. Similarly, because of the ad-hoc nature of their interviews with scammers, no one has ever reported the distribution of the time that scammers take, the size of an average call center, or the amount of money that they charge, all of which can be of immediate use for prioritized take-downs.

In contrast with the aforementioned studies, our work is the first systematic, quantitative study investigating technical support scams, by i) designing and deploying a distributed crawling infrastructure for an 8-month period, ii) using this infrastructure to identify thousands of domains and phone numbers and analyzing their underlying infrastructure, and iii) conducting a controlled, IRB-approved experiment to obtain \emph{precise} information about the social engineering techniques used by scammers and statistics about the process, the tools used, the call-center infrastructures, and the amounts charged.

Though we are not aware of other work that has investigated technical support scams, we argue that these scams are a cross-over between traditional \emph{scareware}, and scams perpetrated over the telephone~\cite{tu16:sok-robocalls} instead of over the Internet, such as \emph{vishing} (Voice Phishing). In a more general sense, our paper belongs to the literature studying cybercrime and underground ecosystems aiming to shine light on hidden mechanisms, affiliate structures, infrastructure abuse, and possible technical and economical disruptions~\cite{park2014scambaiter,shuang2015mules,
motoyama2010re,clark2013there,Kanich:2008:SEA:1455770.1455774,soska2015measuring}.

\vspace{0.5ex}
\noindent\textbf{Scareware.} Scareware refers to software, typically fake AVs, which attempt to scare the user into performing one or more harmful actions. 
Cova et al.~\cite{cova2010analysis} tracked 6,500 domains involved in the distribution of fake AVs and discovered that 65\% of the web servers behind these domains were exclusively serving malicious content. The authors clustered multiple fake AVs as part of the same campaign, with the largest campaign being responsible for 23.5\% of the 6,500 tracked domains. Rajab et al.~\cite{rajab2010nocebo} use Google's SafeBrowsing data to discover over 11,000 domains offering fake AVs with up to 90\% of the discovered scams relying on social engineering for getting installed on a user's computer. 
Stone-Gross et al.~\cite{stone2013underground} approach the phenomenon of fake antivirus software from an economic angle. The authors show that fake AV scammers can earn hundreds of millions of dollars in antivirus license fees and discover the presence of affiliate networks where scammers are paid a commission for each fake AV installation. 
Dietrich et al.~\cite{dietrich2013exploiting} describe how perceptual hashing could be used to automatically cluster malware that depend on visual interfaces including fake antivirus programs and ransomware~\cite{kharrazcutting}.

\vspace{0.5ex}
\noindent\textbf{Telephone Scams.} Maggi performed the first study of vishing by analyzing the data submitted by 360 users who had fallen pray to vishing attacks and were willing to recount their experience~\cite{maggi2010artists}. 
In a later study, Costin et al.~\cite{costinrole} investigated the role of phone numbers in cybercrime and used phone numbers to cluster different types of scams, using data from another crowdsourced website listing scams. The authors utilized HRL (Home Register Location) queries and showed that the average scammer kept, almost always, their phone online. Unfortunately, HRL queries are only applicable to mobile phones, thus we cannot utilize them for tracking toll-free numbers.
Christin et al.~\cite{christin2010dissecting} analyzed a type of scam that was mostly targeting Japanese users by threatening to reveal their adult browsing habits if they would not pay a certain amount of money to scammers. Among others, the authors took advantage of the phone numbers made available by scammers in order to cluster multiple scams as part of larger campaigns. Note that in all three studies, the authors used publicly available data to perform their analyses. Contrastingly, in this paper, because of the absence of available datasets, we designed and developed \tool, the first tool able to automatically discover hundreds of instances of technical support scams on a weekly basis.

Gupta et al. described the architecture of a phone honeypot and presented the intelligence gathered by deploying 40K phone numbers which attracted 1.3 million calls over a period of seven weeks~\cite{gupta2015phoneypot}. The authors discovered that older phone numbers attracted a higher number of calls than newer phone numbers, and showed how the rate of calling can be used to differentiate between different types of unwanted calls, e.g., the ones done by a telemarketer, versus a debt collector. 
While their system could, in principle, be used to discover the older variant of technical support scams (where unsuspecting users receive unsolicited calls from scammers), the type of technical support scams that we investigated in this paper needs an active component, such as \tool, to actively discover pages and numbers.

\section{Conclusion}

In this paper, we reported on the first systematic investigation of technical support scams. By designing and implementing the first system capable of automatically discovering technical support scams, we collected a corpus of thousands of unique domains and telephone numbers engaged in technical support scams, clustered them in campaigns, and showed that scammers abuse specific browser APIs to make it hard for users to navigate away from a technical support scam page. By interacting with 60 different scammers for more than 22 hours, we precisely identified the social engineering techniques used, the remote administration tools abused, and the amount of money that scammers are charging. We presented evidence that places technical support scammers in call centers in English-speaking countries with low wages, showing that the ecosystem of technical support scams is complex and comprised of more than one parties. Lastly, we discussed the need for user education and proposed a simple feature that browser vendors could adopt to assist users in navigating away from malicious pages.

\noindent\textbf{Acknowledgments:}\\
We thank the reviewers for their valuable feedback and Linode for providing us with virtual machines that made our large-scale experiments possible. This work was supported by the Office of Naval Research (ONR) under grant
N00014-16-1-2264, by the National Science Foundation (NSF) under grants CNS-1617902 and CNS-1617593, and by the Cyber Research Institute in Rome, New York.

{\small\bibliographystyle{IEEEtran}
\bstctlcite{IEEEexample:BSTcontrol}
\bibliography{scams}  }

\end{document}